\pdfoutput=1
\documentclass[aps,pra,twocolumn,superscriptaddress,floatfix,10pt]{revtex4-2}

\usepackage{amsmath,amssymb,amsfonts,mathtools,bm,graphicx}
\usepackage[colorlinks=true,allcolors=blue!60!black]{hyperref}
\usepackage{orcidlink}

\newcommand{\vNodesTwoQ}{7}
\newcommand{\vModesTwoQ}{6}
\newcommand{\vModesFourMode}{4}
\newcommand{\vDimDFourMode}{14}
\newcommand{\vDimDFourModeTop}{16}
\newcommand{\vSquidFluxFourMode}{0.32}
\newcommand{\vChiFourMode}{24}
\newcommand{\vTnVsExactFourMode}{below \ensuremath{10^{-3}}~MHz}
\newcommand{\vFourModeSixteenVsFourteenLowFour}{0.22~MHz}
\newcommand{\vNodesFourMode}{5}
\newcommand{\vEjQubitFourMode}{4.5~GHz}
\newcommand{\vEjCouplerFourMode}{1.75~GHz}
\newcommand{\vLQubitFourMode}{95.6~nH}
\newcommand{\vLCouplerLegFourMode}{7.86~nH}
\newcommand{\vCgroundFourMode}{76.1, 38.6, 44.5, 60.4, 133~fF}
\newcommand{\vCcenterFourMode}{10.2, 11.7, 11.5, 9.70~fF}
\newcommand{\vCneighborFourMode}{0.419, 0.673, 0.826~fF}
\newcommand{\vCdistantFourMode}{0.0354, 0.0309, 0.0172~fF}
\newcommand{\vLoopQubitFourMode}{103~nH}
\newcommand{\vElQubitFourMode}{1.58~GHz}
\newcommand{\vLoopCouplerFourMode}{15.7~nH}
\newcommand{\vElCouplerFourMode}{10.4~GHz}
\newcommand{\vNodesThreeQ}{12}
\newcommand{\vModesThreeQ}{11}
\newcommand{\vHilbertThreeQProdApprox}{\ensuremath{10^{9}}}
\newcommand{\vDimDLowThreeQ}{12}
\newcommand{\vDimDHighThreeQ}{4}
\newcommand{\vDimDRefThreeQ}{14}
\newcommand{\vDimDMidProbeThreeQ}{6}
\newcommand{\vDimDStiffProbeThreeQ}{5}
\newcommand{\vDimDAnchorThreeQ}{2}
\newcommand{\vHilbertThreeQStiffProbe}{\ensuremath{2.5\times 10^{9}}}
\newcommand{\vHilbertTwoQFourteenApprox}{\ensuremath{7.5\times 10^{6}}}

\newcommand{\vImplParityTwoQ}{below \ensuremath{10^{-5}}~MHz}
\newcommand{\vTnVsDenseEd}{below \ensuremath{10^{-3}}~MHz}
\newcommand{\vDimDDenseAnchorHigh}{5}
\newcommand{\vTnVsDenseEdEZero}{below \ensuremath{10^{-7}}~MHz}
\newcommand{\vImplParityLevels}{ten}
\newcommand{\vTnVsExactHdFourteen}{0.75~MHz}
\newcommand{\vTnVsExactHdFourteenGround}{0.67~MHz}
\newcommand{\vTnEZeroOffsetFourteen}{0.29~MHz}
\newcommand{\vTnResidualOtherPoints}{0.21~MHz}
\newcommand{\vTnProbeSweeps}{60}
\newcommand{\vTnProbeKrylovDim}{24}
\newcommand{\vTnProbeResidual}{below \ensuremath{10^{-6}}~MHz}
\newcommand{\vDimDCertTwoQ}{14}
\newcommand{\vFourModeSixteenVsFourteen}{2.0~MHz}
\newcommand{\vMpoVsDenseTwoQSweeps}{below \ensuremath{10^{-8}}~MHz}
\newcommand{\vDimDDenseAnchorLow}{4}
\newcommand{\vGridPointsTwoQ}{six}
\newcommand{\vBenchThreads}{16}
\newcommand{\vFockFloorBand}{0.626 to 0.944~MHz}
\newcommand{\vFockFloorTopDim}{18}

\newcommand{\vTolMHz}{1~MHz}
\newcommand{\vTolRelQubit}{0.1\%}
\newcommand{\vTopFockCrit}{\ensuremath{10^{-4}}}
\newcommand{\vChiThreeQ}{32}
\newcommand{\vChiThreeQLowFour}{32}
\newcommand{\vExactMpsBondThreeQ}{20\,736}
\newcommand{\vMpsNumbersThreeQ}{\ensuremath{5.7\times 10^{4}}}
\newcommand{\vBondMoveThreeQ}{10~kHz}
\newcommand{\vBondMoveThreeQLowFour}{0.89~kHz}
\newcommand{\vThreeQBondStepFallFirst}{53}
\newcommand{\vThreeQBondStepFallSecond}{24}
\newcommand{\vTopFockThreeQ}{\ensuremath{4.3\times 10^{-5}}}
\newcommand{\vTopFockStiffProbe}{\ensuremath{1.3\times 10^{-4}}}
\newcommand{\vReplicaSpreadThreeQ}{below 0.1~kHz}
\newcommand{\vReplicaSpreadThreeQLowFour}{below \ensuremath{10^{-5}}~kHz}
\newcommand{\vReplicaCount}{3}
\newcommand{\vThreeQMinLevelSpacing}{110~MHz}
\newcommand{\vBondErrAThirtyTwo}{260~kHz}
\newcommand{\vBondErrLowFourThirtyTwo}{22~kHz}
\newcommand{\vFockErrAThree}{160~kHz}
\newcommand{\vFockErrAThreeLowFour}{15~kHz}
\newcommand{\vTwoQFockMissCount}{three}

\newcommand{\vTnWallEight}{12~min}
\newcommand{\vTnWallSixtyFour}{87~min}
\newcommand{\vTnWallEightSixPoint}{16~min}
\newcommand{\vTnWallSixtyFourSixPoint}{120~min}
\newcommand{\vHdWallExactSixPoint}{81~min}
\newcommand{\vTnMemEight}{0.66~GB}
\newcommand{\vTnMemSixtyFour}{2.1~GB}
\newcommand{\vTnErrSixteen}{0.933~MHz}
\newcommand{\vHdWallExact}{68~min}
\newcommand{\vHdErrFourteen}{below \ensuremath{10^{-6}}~MHz}
\newcommand{\vHdExactMem}{16~GB}
\newcommand{\vEdKrylovVectorsTwoQ}{40}
\newcommand{\vKrylovTolTwoQ}{10~kHz}
\newcommand{\vHdBytesPerState}{2.2~kB}
\newcommand{\vSpeedupBareWall}{5.5}
\newcommand{\vSpeedupBareMem}{25}
\newcommand{\vSpeedupProdWall}{78\%}
\newcommand{\vSpeedupProdMem}{7.9}
\newcommand{\vDressedHubDim}{8}

\newcommand{\vThreeQWallSixteen}{27~min}
\newcommand{\vThreeQWallThirtyTwo}{81~min}
\newcommand{\vThreeQWallSixtyFour}{4.4~h}
\newcommand{\vThreeQWallOneTwentyEight}{13~h}
\newcommand{\vThreeQMemSixteen}{0.97~GB}
\newcommand{\vThreeQMemThirtyTwo}{1.3~GB}
\newcommand{\vThreeQMemSixtyFour}{2.4~GB}
\newcommand{\vThreeQMemOneTwentyEight}{5.8~GB}
\newcommand{\vNodeMem}{230~GB}
\newcommand{\vCpuModel}{Intel Xeon 6517P}
\newcommand{\vJuliaVersion}{1.11.6}
\newcommand{\vITensorsVersion}{0.7.13}
\newcommand{\vITensorMPSVersion}{0.3.6}
\newcommand{\vKrylovKitVersion}{0.9.5}
\newcommand{\vEdDenseMemThreeQ}{\ensuremath{1.7\times 10^{10}}~GB}
\newcommand{\vHdElevenModeMem}{2.2~TB}
\newcommand{\vHdGuideHours}{66~h}
\newcommand{\vCompileTnFourMode}{\ensuremath{40\pm 20}~s}
\newcommand{\vCompileHdFourMode}{\ensuremath{1.65\pm 0.04}~s}

\newcommand{\vSweeps}{60}
\newcommand{\vSweepsTwoQ}{40}
\newcommand{\vMaxdimRamp}{64, 128, 256, 512, 800, 1000}
\newcommand{\vCutoff}{\ensuremath{10^{-13}}}
\newcommand{\vNoiseRamp}{\ensuremath{10^{-4}}, \ensuremath{10^{-6}}, \ensuremath{10^{-8}}, \ensuremath{10^{-10}}, 0}
\newcommand{\vKrylovDim}{12}

\newcommand{\vEffModesThreeQ}{1.7 to 4.2}
\newcommand{\vEffModesThreeQQubitOne}{2.9}
\newcommand{\vEffModesThreeQQubitThree}{4.2}
\newcommand{\vEffModesTwoQQubit}{2.0}
\newcommand{\vEffModesTwoQCoupler}{1.1}
\newcommand{\vTopModeWeightWorstThreeQ}{0.30}
\newcommand{\vDoubletFreqTwoQ}{6.805~GHz}
\newcommand{\vFZeroOneTwoQZThirty}{1.076~GHz}

\newcommand{\vTnLadderTotal}{20~h}
\newcommand{\vTnLadderCaps}{4.1~h}
\newcommand{\vTnLadderFock}{9.8~h}
\newcommand{\vTnLadderReplicas}{5.9~h}
\newcommand{\vHdLadderTotal}{2.0~h}
\newcommand{\vLadderRatio}{10\%}
\newcommand{\vChiTwoQ}{4 to 6}
\newcommand{\vChiTwoQLowFour}{4 to 6}
\newcommand{\vChiGrowthRange}{about three to eight}

\newcommand{\vThreeQLadderHours}{19~h}
\newcommand{\vThreeQFockSweepSolves}{12}
\newcommand{\vThreeQFockSweepMaxMem}{3.0~GB}
\newcommand{\vFockOffPathStiff}{630~kHz}
\newcommand{\vFockOffPathStiffLowFour}{630~kHz}
\newcommand{\vFockOffPathMid}{0.47~kHz}
\newcommand{\vFockOffPathMidLowFour}{0.19~kHz}
\newcommand{\vThreeQBondMoveMax}{33~kHz}
\newcommand{\vThreeQBondMoveProd}{0.50~kHz}
\newcommand{\vThreeQWallGrowth}{28}
\newcommand{\vThreeQMemGrowth}{5.9}

\newcommand{\vDimDBareRefThreeQ}{10}

\newcommand{\vCshunt}{10~fF}
\newcommand{\vCgroundTwoQ}{50~fF}
\newcommand{\vLFourThreeQ}{14.3~nH}
\newcommand{\vLCouplerLegThreeQ}{8.17~nH}
\newcommand{\vLEndThreeQ}{59.3~nH}
\newcommand{\vLBridgeThreeQ}{36.8~nH}
\newcommand{\vCqubitThreeQ}{70~fF}
\newcommand{\vCcouplerThreeQ}{25~fF}
\newcommand{\vChubThreeQ}{25~fF}
\newcommand{\vEcQubitThreeQ}{0.430~GHz}
\newcommand{\vFluxQuantumConst}{163~GHz\,nH}
\newcommand{\vEjQubitTwoQ}{8.98~GHz}
\newcommand{\vEjCouplerTwoQ}{3.5~GHz}
\newcommand{\vLoopQubitTwoQ}{108~nH}
\newcommand{\vElQubitTwoQ}{1.52~GHz}
\newcommand{\vEjElQubitTwoQ}{5.92}
\newcommand{\vLoopCouplerTwoQ}{15.6~nH}
\newcommand{\vElCouplerTwoQ}{10.5~GHz}
\newcommand{\vEjElCouplerTwoQ}{0.334}
\newcommand{\vEcQubitTwoQ}{0.553~GHz}
\newcommand{\vEjQubitThreeQ}{3.70~GHz}
\newcommand{\vEjCouplerThreeQ}{10~GHz}
\newcommand{\vLoopQubitThreeQ}{81.7~nH}
\newcommand{\vElQubitThreeQ}{2.00~GHz}
\newcommand{\vEjElQubitThreeQ}{1.85}
\newcommand{\vLoopCouplerThreeQ}{16.3~nH}
\newcommand{\vElCouplerThreeQ}{10.0~GHz}
\newcommand{\vEjElCouplerThreeQ}{1.00}
\newcommand{\vBetaLoadedMinThreeQ}{0.970}
\newcommand{\vBetaLoadedMaxThreeQ}{0.982}
\newcommand{\vBetaLoadedMinHalfFluxThreeQ}{0.949}
\newcommand{\vBetaLoadedMaxHalfFluxThreeQ}{0.978}
\newcommand{\vMinimaThreeQ}{8}
\newcommand{\vMinimaMaxThreeQ}{8}
\newcommand{\vLBodyTwoQ}{50~nH}
\newcommand{\vLCouplerLegTwoQ}{7.8~nH}
\newcommand{\vFZeroOneThreeQ}{750~MHz}
\newcommand{\vLocalOscTwoQ}{3.063, 4.839, 2.607, 2.598, 8.734, 14.576}
\newcommand{\vModeFreqTwoQ}{3.088, 4.840, 6.804, 6.807, 9.903, 14.581}
\newcommand{\vLocalOscThreeQLow}{1.929, 2.985, 2.622, 4.049, 3.977}
\newcommand{\vLocalOscThreeQMid}{8.185, 8.963}
\newcommand{\vLocalOscThreeQStiff}{13.254, 13.455, 21.112, 21.536}
\newcommand{\vModeFreqThreeQLow}{2.267, 3.684, 4.270, 4.934, 5.235}
\newcommand{\vModeFreqThreeQMid}{8.380, 9.151}
\newcommand{\vModeFreqThreeQStiff}{17.173, 17.350, 21.166, 21.558}
\newcommand{\vModeSpanThreeQ}{2.267 to 21.558~GHz}
\newcommand{\vDoubletPencilPair}{6.804 and 6.807~GHz}
\newcommand{\vDoubletLocalPair}{2.607 and 2.598~GHz}
\newcommand{\vDoubletConventionRatio}{2.6}

\newcommand{\vWarmColdDevMax}{0.097~MHz}
\newcommand{\vTwoQFockMissZThirtyA}{1.6~MHz}
\newcommand{\vTwoQFockMissZThirtyB}{1.04~MHz}
\newcommand{\vTwoQTopFock}{below \ensuremath{10^{-4}}}
\newcommand{\vThreeQBridgeMax}{below \ensuremath{10^{-9}}~MHz}
\newcommand{\vMaxdimRampTwoQ}{16, 32, then the cap}
\newcommand{\vWarmSpeedupCapEight}{1.1}
\newcommand{\vWarmSpeedupCapSixteen}{1.1}
\newcommand{\vWarmSpeedupCapThirtyTwo}{1.2}
\newcommand{\vWarmSpeedupCapSixtyFour}{0.85}
\newcommand{\vWarmStepsPassed}{7 of 20}
\newcommand{\vWarmStepSpeedup}{3.8 to 4.7}
\newcommand{\vThreeQWarmFailedSteps}{2}
\newcommand{\vThreeQWarmReturnMHz}{below \ensuremath{10^{-5}}~MHz}
\newcommand{\vTwoQFockStepFirstPointTwo}{1.1~MHz}
\newcommand{\vTwoQFockStepFirstPointThree}{1.7~MHz}
\newcommand{\vTwoQFockStepSecondPointTwo}{0.57~MHz}
\newcommand{\vTwoQFockStepSecondPointThree}{0.97~MHz}
\newcommand{\vTwoQFockTwentyVsFourteen}{1.01~MHz}
\newcommand{\vTwoQFockSixteenRerun}{0.020~MHz}
\newcommand{\vThreeQTransitionOne}{0.422~GHz}
\newcommand{\vThreeQFockBarOne}{5.2~kHz}
\newcommand{\vThreeQStiffShiftOne}{140~kHz}
\newcommand{\vThreeQBondBarOne}{0.76~kHz}
\newcommand{\vThreeQTransitionTwo}{0.705~GHz}
\newcommand{\vThreeQFockBarTwo}{2.3~kHz}
\newcommand{\vThreeQStiffShiftTwo}{100~kHz}
\newcommand{\vThreeQBondBarTwo}{0.10~kHz}
\newcommand{\vThreeQTransitionThree}{0.816~GHz}
\newcommand{\vThreeQFockBarThree}{13~kHz}
\newcommand{\vThreeQStiffShiftThree}{480~kHz}
\newcommand{\vThreeQBondBarThree}{0.89~kHz}
\newcommand{\vThreeQTransitionFour}{1.038~GHz}
\newcommand{\vThreeQFockBarFour}{43~kHz}
\newcommand{\vThreeQStiffShiftFour}{96~kHz}
\newcommand{\vThreeQBondBarFour}{3.1~kHz}
\newcommand{\vThreeQTransitionFive}{1.220~GHz}
\newcommand{\vThreeQFockBarFive}{2.2~kHz}
\newcommand{\vThreeQStiffShiftFive}{180~kHz}
\newcommand{\vThreeQBondBarFive}{2.0~kHz}
\newcommand{\vThreeQTransitionSix}{1.477~GHz}
\newcommand{\vThreeQFockBarSix}{15~kHz}
\newcommand{\vThreeQStiffShiftSix}{440~kHz}
\newcommand{\vThreeQBondBarSix}{4.5~kHz}
\newcommand{\vThreeQTransitionSeven}{1.771~GHz}
\newcommand{\vThreeQFockBarSeven}{160~kHz}
\newcommand{\vThreeQStiffShiftSeven}{78~kHz}
\newcommand{\vThreeQBondBarSeven}{10~kHz}
\newcommand{\vOrthoWeight}{100}
\newcommand{\vGuardStatesTwoQ}{4}
\newcommand{\vGuardStatesThreeQ}{3}
\newcommand{\vInitBondTwoQ}{20}
\newcommand{\vInitBondThreeQ}{30}
\newcommand{\vWarmSweeps}{10}
\newcommand{\vWarmNoise}{\ensuremath{10^{-8}}}
\newcommand{\vWarmOrthoTol}{\ensuremath{10^{-3}}}
\newcommand{\vWarmVarianceFactor}{10}
\newcommand{\vWarmSettleTol}{0.01~MHz}
\newcommand{\vColdResolveSpread}{0.75 to 1.2}
\newcommand{\vWarmVsRefSeeded}{0.11~MHz}
\newcommand{\vWarmVsRefCold}{0.029 to 0.043~MHz}
\newcommand{\vWarmStepsFailed}{13}
\newcommand{\vWarmCrossingOnly}{5}

\begin{document}

\title{Tensor-network simulation of strongly coupled multi-fluxonium circuits}

\author{Christian Hensel\,\orcidlink{0000-0002-0319-7908}}
\email{chris.hensel@qilimanjaro.tech}
\affiliation{Qilimanjaro Quantum Tech, Carrer de Vene\c{c}uela 74, 08019 Barcelona, Spain}
\affiliation{Universitat Autònoma de Barcelona, 08193 Bellaterra (Barcelona), Spain}

\author{Andrés N. Cáliz\,\orcidlink{0009-0002-6938-8595}}
\email{andres.navas@qilimanjaro.tech}
\affiliation{Qilimanjaro Quantum Tech, Carrer de Vene\c{c}uela 74, 08019 Barcelona, Spain}
\affiliation{Departament de Física, Universitat de Barcelona, 08007 Barcelona, Spain}

\author{Pau Torrente\,\orcidlink{0009-0005-2942-3263}}
\affiliation{Qilimanjaro Quantum Tech, Carrer de Vene\c{c}uela 74, 08019 Barcelona, Spain}
\affiliation{Barcelona Supercomputing Center, Plaça Eusebi Güell, 1-3, 08034 Barcelona, Spain}

\author{María Hita-Pérez\,\orcidlink{0000-0002-8330-6237}}
\affiliation{Qilimanjaro Quantum Tech, Carrer de Vene\c{c}uela 74, 08019 Barcelona, Spain}

\author{Arnau Riera\,\orcidlink{0000-0002-3271-7802}}
\affiliation{Qilimanjaro Quantum Tech, Carrer de Vene\c{c}uela 74, 08019 Barcelona, Spain}

\date{\today}

\begin{abstract}
Designing a superconducting quantum processor rests on predicting the
low-lying spectrum of its coupled circuit. The Hilbert-space dimension of the
circuit's lumped-element model grows exponentially with the number of circuit
nodes. The usual
methods split the circuit into subsystems, but that split is hard to choose
when the elements are strongly coupled. We present a tensor-network method
that chooses no subsystems and whose memory and run time per solve are set by
the number of modes and two truncations, the Fock cutoff of each normal mode and the bond
dimension of a matrix product state (MPS), not by the Hilbert-space dimension. From
the capacitance matrix, the inverse-inductance matrix and the junction list
alone, the linear circuit is diagonalized into normal modes. Each junction
cosine is then a sum of two products of single-mode operators, so the complete
lumped-element Hamiltonian is a matrix product operator (MPO) of small bond
dimension with one site per mode, solved by the density matrix renormalization
group (DMRG). The normal modes make the method generic, and the tensor network
makes it affordable. For two fluxonia joined by a tunable coupler, the method agrees with
exact diagonalization of the same truncated space to within \vTolMHz{}. Our
circuit of three fluxonia joined by two couplers has \vModesThreeQ{} normal
modes and a truncated space of \vHilbertThreeQProdApprox{} states, out of reach
of exact diagonalization in dense or sparse storage. At a bond dimension of
\vChiThreeQ{}, its lowest eight levels are converged to within \vTolMHz{} in
both truncations, and each state is stored in a small fraction of the numbers
a dense vector needs. One solve at that bond dimension takes
\vThreeQWallThirtyTwo{} and \vThreeQMemThirtyTwo{} of memory on one CPU node.

\end{abstract}

\maketitle

\section{Introduction}\label{sec:intro}

Superconducting circuits are a leading platform for quantum information
processing~\cite{kjaergaard_superconducting_2020,krantz_quantum_2019}, including adiabatic and analog
processors~\cite{albash_adiabatic_2018,johnson_quantum_2011,king_quantum_2023}.
What such a processor computes is set by the effective Hamiltonian of its
coupled circuit elements, so the design process rests on predicting the
low-lying spectrum of a circuit. In the circuits
studied here the elements are deliberately strongly coupled (defined below),
which makes that prediction expensive.

The Hilbert-space dimension of a lumped-element model grows exponentially with
the number of circuit nodes. Exact diagonalization of the complete model is
still affordable for the two coupled fluxonia studied in this paper, but out of
reach as a method for three.
The standard responses split the circuit into subsystems. Born-Oppenheimer separations of fast and
slow modes~\cite{Kafri2017}, and hierarchical diagonalization schemes that
diagonalize subsystems and keep a fixed number of levels per
subsystem~\cite{Kerman2020,groszkowski_scqubits_2021,chitta_computer-aided_2022},
both make the problem tractable. A Born-Oppenheimer separation holds when the modes it eliminates are
much faster than the ones it keeps, and hierarchical diagonalization agrees
with exact results once its kept-level ladder, the sequence of solves at an
increasing number of kept levels per subsystem, has converged. But each of these tools starts from a
decomposition of the circuit into subsystems that the user chooses, a design
step repeated for every circuit. A poor partition converges slowly in the number
of kept levels, and an apparent plateau of the ladder tests the number of kept
levels but not the partition itself or the basis in which each subsystem
was diagonalized. A wrong choice can therefore pass the standard workflow
undetected. Methods built on such a split have documented
limits of this kind: for two coupled flux qubits, perturbative estimates of
the qubit-qubit couplings agree with a nonperturbative projection onto the
low-energy subspace only at weak coupling~\cite{hita-perez_three-josephson_2021}.

Matrix product
operator (MPO) representations solved with the density matrix renormalization
group
(DMRG)~\cite{orus_practical_2014,orus_tensor_2019,verstraete_matrix_2008,white_density_1992,schollwoeck_density-matrix_2011}
have been applied to superconducting circuits before, with one site per
superconducting island~\cite{chung_superconductor_1997,lee_quantum_2003}, per
junction of an array
inductance~\cite{dipaolo_efficient_2021,rui_tensor-network_2025} or per
qubit and per
coupler~\cite{gonzalez-garcia_multi-target_2025,lange_cross-talk_2025}.
Reference~\cite{weiss_spectrum_2019} applies DMRG to an effective model of the
current-mirror qubit. We instead
diagonalize the linear part of the whole circuit into normal
modes~\cite{ferguson_symmetries_2013,smith_quantization_2016} and use those
modes as the sites. The input is the capacitance matrix $C$, the geometric inverse-inductance matrix $L^{-1}$
and the list of Josephson junctions, and these fix the mode basis: no subsystem
is identified and no junction is assigned to a local mode. Each mode carries a
truncated Fock space, the junction cosines are kept in full and spread over the
modes they thread, and the complete lumped-element Hamiltonian, linear
inductances included, becomes an MPO with one site per mode. DMRG then returns
the low-lying spectrum under two truncations only, a per-mode Fock cutoff and
the bond dimension, both of which are swept. The normal modes are not local,
since every junction acts on several of them. The hierarchical solver of
Ref.~\cite{Kerman2020} avoids them because their couplings to the junctions
are dense operators that slow its diagonalization.
In the tensor network, the cost of a solve is set by the number of modes, the
bond dimension and the per-site dimension
(Sec.~\ref{sec:q3-cost}), so both truncations can be swept where the Hilbert
space is far too large to store.

We call these circuits \emph{strongly coupled} in a structural sense. The
usual sense is a large coupling between two chosen
subsystems~\cite{blais_circuit_2021}. Here we mean that the normal modes
hybridize the junctions, so that no junction of the three-fluxonium circuit
belongs to a single normal mode. By the participation ratio of Appendix~\ref{app:modes}, its
junctions span \vEffModesThreeQ{} modes, so no junction cosine can be folded
into the local basis of one site, and each acts on several sites at once. At
two fluxonia the coupler junction sits almost on one mode, at
\vEffModesTwoQCoupler, and the qubit junctions at \vEffModesTwoQQubit, a value
set by the exchange symmetry of two identical qubits. The mode basis leaves
the flux bias out, since every junction is linearized about zero phase
(Sec.~\ref{sec:mpo-modes}). The participation ratio is therefore the same at
every bias and describes the basis, while the qubit-qubit interaction changes
with the coupler bias.

On two fluxonia joined by a tunable coupler, where exact diagonalization of the
same truncated space is still affordable, the tensor network agrees with it to
within \vTolMHz{} (Sec.~\ref{sec:q2}).
The circuit of three fluxonia with two couplers has \vModesThreeQ{} normal
modes and a space of \vHilbertThreeQProdApprox{} states. With the same method
and implementation, its lowest eight levels at one bias point meet every
criterion of Appendix~\ref{app:convergence} at
one setting. That setting has bond dimension \vChiThreeQ, and one solve there
takes \vThreeQWallThirtyTwo{} and \vThreeQMemThirtyTwo{} on one CPU node
(Sec.~\ref{sec:q3}).
Exact diagonalization in the same basis is out of reach there as a method. A
dense matrix would occupy \vEdDenseMemThreeQ{}. Sparse storage, which keeps
only the nonzero entries, would still exceed the memory of the CPU node, since
every junction cosine acts on all eleven modes. A matrix-free exact solve,
one that applies the Hamiltonian to vectors without storing it as a matrix,
was not attempted at that size (Sec.~\ref{sec:q3-cost}). A dressed local basis
built on the normal modes would first assign each junction to one mode and build that mode's basis from
its own oscillator and the junction cosines assigned to it. At eleven modes no
junction sits on one mode, so that assignment has no natural answer, and the
normal-mode basis needs none.

The rest of the paper is organized as follows. Section~\ref{sec:method}
constructs the normal-mode MPO and its two truncations, and
Sec.~\ref{sec:circuits} introduces the two circuits. Section~\ref{sec:q2}
validates the tensor network against exact diagonalization and measures the
cost of both, Sec.~\ref{sec:q3} takes the method to eleven modes, and
Sec.~\ref{sec:disc} discusses the results and the questions they leave open.
Appendix~\ref{app:params} lists the circuit parameters,
Appendix~\ref{app:modes} details the normal modes and the participation of
each junction, Appendix~\ref{app:convergence} states the convergence
criteria and the DMRG schedule, and Appendix~\ref{app:hardware} describes the
CPU node and the software.

\section{Normal-mode matrix product operators}\label{sec:method}

We write one lumped-element Hamiltonian for the whole circuit, qubits and
couplers together, so no weak-interaction assumption enters. We then map it to
an MPO with one site per normal mode and define the two truncations under which
it is solved.

\subsection{Lumped-element Hamiltonian}

In terms of the node charges $\mathbf{Q}$ and node fluxes $\bm{\Phi}$ of the
circuit graph~\cite{Vool2017, GarciaRipoll2022}, the Hamiltonian reads
\begin{equation}
\begin{split}
H(\mathbf{Q},\bm{\Phi}) ={}&
\frac{1}{2}\,\mathbf{Q}^{\mathrm{T}} C^{-1}\mathbf{Q}
+ \frac{1}{2}\,\bm{\Phi}^{\mathrm{T}} L^{-1} \bm{\Phi}\\
&- \sum_{b\in\mathrm{JJ}} E_{J,b}\cos\!\left(\phi_b + \theta_b\right),
\end{split}
\label{eq:lumped}
\end{equation}
with $C$ the capacitance matrix, $L^{-1}$ the geometric inverse-inductance
matrix and $E_{J,b}$ the Josephson energy of junction $b$, which joins nodes
$i(b)$ and $j(b)$. The phase across it is $\phi_b = \phi_{i(b)} - \phi_{j(b)}$,
where $\phi_i = \Phi_i/\varphi_0$ is the phase of node $i$ and
$\varphi_0=\Phi_0/2\pi$ the reduced flux quantum. Each junction closes one flux
loop (Appendix~\ref{app:params}), and the external flux $\Phi_\ell$
threading that loop enters only through the phase offset
$\theta_b = 2\pi\Phi_\ell/\Phi_0$ of the junction's Josephson term. Fluxes are quoted in units of
$\Phi_0$ with the loop labels of the full circuit diagrams in
Appendix~\ref{app:params}. Unless noted otherwise, every qubit loop
sits at $\Phi_q = \Phi_0/2$. A coupler loop is unbiased at $\Phi_c = 0$ and
frustrated at $\Phi_c = \Phi_0/2$.

\subsection{Normal-mode basis}
\label{sec:mpo-modes}

A discrete charge basis is natural for an island joined to the circuit by
capacitors and junctions alone, but it encodes an inductively shunted node
inefficiently. The inductance makes the node phase an extended variable,
so its conjugate charge becomes continuous, and a charge basis
resolves it only on a fine grid with a large cutoff. Fock states of an
oscillator suit an extended phase. We therefore diagonalize the linear part of
the circuit first and let each normal mode carry its own truncated Fock
space~\cite{nigg_black-box_2012, minev_energy-participation_2021, smith_quantization_2016, rajabzadeh_analysis_2023}. The
input is $C$, $L^{-1}$ and the junction list.

The linearized circuit is used only to choose the mode coordinates. Linearizing every junction about zero phase adds
$(2\pi/\Phi_0)^2 E_{J,b}$ to the inverse-inductance matrix between its two
nodes, giving a total $L^{-1}_{\mathrm{tot}}$. Together with $C$ it fixes the
normal modes of the linearized circuit. The added term does not depend on the
offset $\theta_b$, so $L^{-1}_{\mathrm{tot}}$ and the mode basis are the same at
every bias. The Hamiltonian solved is still Eq.~\eqref{eq:lumped}, since the
change to mode coordinates is linear and every junction keeps its full cosine,
$\theta_b$ included. The inductive matrix $E_L$ of
Eq.~\eqref{eq:mode-matrices} is therefore built from the geometric $L^{-1}$
alone, because the quadratic part of each junction is already inside its
cosine. Solving
the generalized eigenproblem
\begin{equation}
L^{-1}_{\mathrm{tot}}\, v_\mu = \omega_\mu^2 \, C \, v_\mu
\label{eq:normalmodes}
\end{equation}
yields the mode transformation $M$, whose columns are the eigenvectors $v_\mu$
and which relates the node phases of Eq.~\eqref{eq:lumped} to mode
coordinates, $\phi_i = \sum_\mu M_{i\mu}\,\xi_\mu$. The eigenvectors are
$C$-orthogonal, and each is scaled so that its largest node entry has
magnitude one. When the inductors and junctions join all $N$ nodes and only
capacitances lead to ground, the inductive network has zero row sums.
Equation~\eqref{eq:normalmodes} then carries one zero eigenvalue, the global phase.
Its conjugate, the total charge, is conserved, so dropping it only fixes the
charge sector and leaves $N-1$ modes. In the mode
basis the charging and inductive matrices become
\begin{equation}
\begin{split}
E_{C,\mu\nu} &= \frac{e^2}{2}\big(M^{\mathrm{T}} C M\big)^{-1}_{\mu\nu},\\
E_{L,\mu\nu} &= \varphi_0^{2}\,\big(M^{\mathrm{T}} L^{-1} M\big)_{\mu\nu},
\end{split}
\label{eq:mode-matrices}
\end{equation}
and each junction acquires a participation vector
\begin{equation}
\alpha^{(b)}_\mu = M_{i(b),\mu} - M_{j(b),\mu},
\end{equation}
recording how strongly mode $\mu$ threads the branch between the junction's
two nodes $i(b)$ and $j(b)$. A junction thus collects phase from every mode
that moves its two nodes differently.

The MPO has one site per mode, with local dimension equal to that mode's Fock
cutoff, and three kinds of terms. The diagonal part is a local harmonic
Hamiltonian,
\begin{equation}
H^{\mathrm{loc}}_\mu = \Omega_\mu\left(a_\mu^{\dagger}a_\mu + \tfrac12\right),
\qquad \Omega_\mu = 2\sqrt{2 E_{C,\mu\mu} E_{L,\mu\mu}},
\label{eq:local}
\end{equation}
with
$\hat\phi_\mu = (8E_{C,\mu\mu}/E_{L,\mu\mu})^{1/4}(a_\mu+a^{\dagger}_\mu)/\sqrt2$
and
$\hat n_\mu = (E_{L,\mu\mu}/8E_{C,\mu\mu})^{1/4}(a_\mu-a^{\dagger}_\mu)/(i\sqrt2)$.
Here $\Omega_\mu$, the frequency of the oscillator whose Fock states site
$\mu$ is built on, differs from $\omega_\mu$ of Eq.~\eqref{eq:normalmodes}
because $E_L$ holds the geometric inductance only (Appendix~\ref{app:modes}
lists both). The off-diagonal entries
of $E_L$ give two-site terms $E_{L,\mu\nu}\hat\phi_\mu\hat\phi_\nu$. $E_C$ has
none, because the columns of $M$ are $C$-orthogonal.

Each junction cosine spreads
across all modes through its participations,
\begin{equation}
\begin{split}
-E_{J,b}&\cos\!\Big(\textstyle\sum_\mu \alpha^{(b)}_\mu \hat\phi_\mu + \theta_b\Big)\\
&= -\frac{E_{J,b}}{2}\Big( e^{\,i\theta_b}\!\prod_\mu
   e^{\,i\alpha^{(b)}_\mu\hat\phi_\mu} + \mathrm{h.c.}\Big),
\end{split}
\label{eq:junction-product}
\end{equation}
where $\theta_b$ is the phase offset of Eq.~\eqref{eq:lumped}. The right-hand
side of Eq.~\eqref{eq:junction-product} is a \emph{product of single-site
operators}: each factor $\exp(i\alpha^{(b)}_\mu\hat\phi_\mu)$ is a dense
matrix on one site, computed once by exponentiating that mode's truncated
$\hat\phi_\mu$. A junction that couples every mode therefore costs only one
operator string, and the whole Hamiltonian is an
operator sum that \textsc{ITensor}'s \texttt{OpSum} compresses to an MPO of
modest bond dimension~\cite{itensor, chan_matrix_2016}.

The external fluxes enter only through the
scalar coefficients $e^{\pm i\theta_b}$ of Eq.~\eqref{eq:junction-product}. The
per-site matrices thus depend on the mode basis alone, which is fixed for all
biases. A flux
sweep therefore rebuilds only the coefficient list. A state converged at one
bias is also an admissible starting vector at the next, a standard
continuation~\cite{stoudenmire_studying_2012}.

\subsection{DMRG and the two truncation parameters}
\label{sec:truncation}

Low-lying eigenstates are obtained with
DMRG~\cite{white_density_1992, schollwoeck_density-matrix_2011}. The lowest states
of a multi-qubit circuit form a closely spaced cluster whose members must not
collapse onto one another. Excited states are therefore computed one at a time,
each as its own matrix product state. An energy penalty for overlap with the
states already found keeps each new one orthogonal to them.
Appendix~\ref{app:convergence} gives the penalty weight and the other solver
settings. Every solve of the
convergence study starts from a random state with a recorded seed. Every
tensor-network solve is built from the public
\textsc{ITensor} package~\cite{itensor}.

The encoding has exactly two truncation parameters, the Fock cutoff of each
mode and the bond cap, the largest bond dimension the matrix product state may
reach. They are set
separately, and both are swept. Every comparison in Secs.~\ref{sec:q2} and~\ref{sec:q3}
uses one metric, the worst deviation from a stated reference of a transition
between any two of the quoted levels, over all bias points of the stated grid.
Section~\ref{sec:q2} also gives the worst deviation among the transitions
from the ground state.

The reference is of two kinds.
\emph{Same-space agreement} compares the tensor network with an exact solve of
the same truncated Hamiltonian: it measures the DMRG error alone and exists
only for two fluxonia (Sec.~\ref{sec:q2}). \emph{Truncation convergence} compares a solve
with the largest setting computed along one truncation axis, Fock or bond, the
other held fixed: its deviation, the Fock or bond error of that solve, measures
how far the transitions still move when that truncation is raised; it exists
for both circuits and is quoted per axis, never summed.

A solve is in the
\vTolMHz{} class when its deviation from the stated reference is below
\vTolMHz. The tolerance is absolute, since couplings and
level shifts are differences of transitions; for scale, \vTolMHz{} is about
\vTolRelQubit{} of the qubit transitions of Sec.~\ref{sec:circuits}
(\vFZeroOneThreeQ{} for an isolated fluxonium of the three-fluxonium circuit,
up to \vFZeroOneTwoQZThirty{} on the two-fluxonium grid). Both references are
themselves truncated, so neither measures how far a solve lies from the
untruncated Hamiltonian. Convergence within \vTolMHz{} says only that the
transitions moved by less than \vTolMHz{} up to the largest setting run, and an
error shared by every setting on an axis does not show up.

DMRG lets the bond dimension grow during its sweeps up to the bond cap, and
each solve is labeled by its cap. The bond dimension $\chi$ quoted for a circuit is the smallest cap from which
every further step up the bond ladder moves the quoted transitions by less
than \vTolMHz. A solve at cap $\chi$ can still reach a smaller bond
dimension. Because $\chi$ is defined against the tolerance, it cannot grow when
the tolerance is loosened, so a looser target is met by cheaper solves.

Solves are repeated from independent random initial states, called replicas,
to test for local minima. The bond axis is variational, so its ground-state energy error is
one-sided. The Fock axis is not variational, so it can only test for a plateau.
Per circuit size, the construction needs only a mode ordering, a per-mode Fock
cutoff and a sweep schedule. Appendix~\ref{app:convergence} records these
choices, states the convergence criteria and lists for each circuit and bias
point which axes meet them.

\section{Two multi-fluxonium circuits}\label{sec:circuits}

We study two circuits, one with two
fluxonia~\cite{manucharyan_fluxonium_2009, nguyen_high-coherence_2019,
somoroff_millisecond_2023, weiss_fast_2022, zhang_universal_2021,
nguyen_blueprint_2022, bao_fluxonium_2022} and one with three, in which
neighboring qubits are joined through flux-tunable
couplers~\cite{yan_tunable_2018, sung_realization_2021, moskalenko_high_2022,
ding_high-fidelity_2023}, here rf-SQUID loops that couple
inductively~\cite{Kafri2017}. Both belong to one family, defined by a
construction rule: the inductors connect every node without closing a loop, so
every junction closes exactly one flux loop, and each qubit loop shares one
inductive leg with each neighboring coupler loop. The only capacitances are one
from each node to ground and a shunt across each qubit junction, so the qubits
couple through the inductive network alone. The element values of both
circuits are design values, not extracted from a layout. In an implementation
every inductance of both circuits is a chain of large Josephson
junctions~\cite{manucharyan_fluxonium_2009}, each with a Josephson energy far
above its charging energy. Phase slips through such junctions are
rare~\cite{masluk_microwave_2012}, so we model each chain as a linear
inductance. This leaves out the chain's internal
modes~\cite{ferguson_symmetries_2013}. A stray capacitance
between qubits, as a fabricated chip would have, would only change the
capacitance matrix $C$, which the construction of Sec.~\ref{sec:method} takes
in full. The family was chosen because it is a building
block of an analog processor. Besides size and layout, the two circuits differ
in operating regime (Appendix~\ref{app:params}). The ratio $E_J/E_L$ of each
qubit loop falls from \vEjElQubitTwoQ{} to \vEjElQubitThreeQ{}, so at half a
flux quantum the qubit potential goes from a deep to a shallow double well. The
same ratio for a coupler loop rises from \vEjElCouplerTwoQ{} to
\vEjElCouplerThreeQ{}.

For an rf SQUID the ratio of a coupler loop is the screening parameter, and a
coupler stays monostable while it is below 1~\cite{Kafri2017}. In the
three-fluxonium circuit the rest of the network loads each coupler
junction~\cite{Kafri2017,Quintana2017}. At the working point of
Sec.~\ref{sec:q3} this lowers each coupler's screening parameter to between
\vBetaLoadedMinThreeQ{} and \vBetaLoadedMaxThreeQ{}
(Appendix~\ref{app:params}), just inside the monostable regime. There a linear
model of the coupler's response breaks down toward half a flux
quantum~\cite{Quintana2017}. The method keeps every junction's full cosine, and
the couplers sit unbiased at the working point, where their potential is
stiffest.

\begin{figure*}[t]
\centering
\makebox[\textwidth][l]{\textbf{(a)}}\\[1mm]
\includegraphics[width=0.59\textwidth]{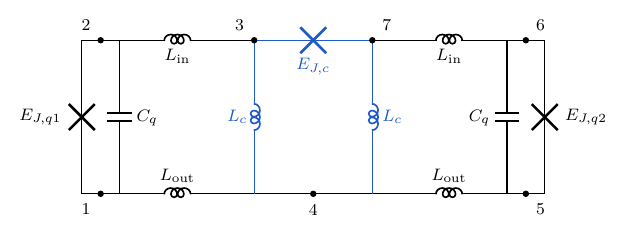}\\[3mm]
\makebox[\textwidth][l]{\textbf{(b)}}\\[1mm]
\includegraphics[width=0.92\textwidth]{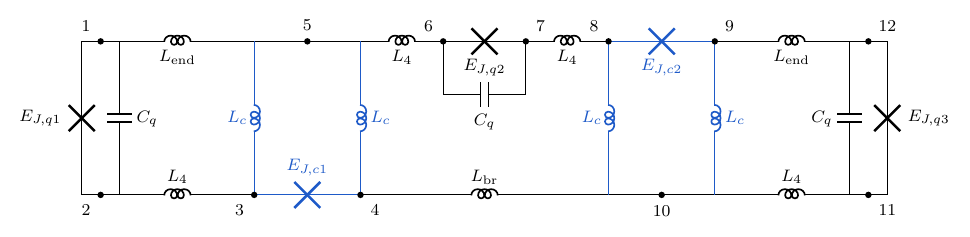}
\caption{(a) The two-fluxonium circuit: \vNodesTwoQ{} nodes, two qubit
junctions on branches $(1,2)$ and $(5,6)$, and one flux-tunable coupler
junction on branch $(3,7)$ whose two legs meet at node 4.
(b) The three-fluxonium circuit: \vNodesThreeQ{} nodes, three qubit junctions on
branches $(1,2)$, $(6,7)$ and $(11,12)$ and two coupler junctions on $(4,3)$
and $(8,9)$, in the sequence q1, c1, q2, c2, q3, with the coupler legs
meeting at nodes 5 and 10. In both panels Josephson junctions are drawn as crosses, and
the coupler junctions and their legs are drawn in blue.
Appendix~\ref{app:params} uses the same node numbers, gives every element
value and draws the ground capacitances and flux loops omitted here.}
\label{fig:circuits}
\end{figure*}

\subsection{Two fluxonia with a tunable coupler}

The two-fluxonium circuit, drawn in Fig.~\ref{fig:circuits}(a), follows the
galvanically coupled fluxonium pair of Ref.~\cite{weiss_fast_2022}. Each
fluxonium's inductance is split into an outer arm $L_{\mathrm{out}}$ to node 4, where the two
coupler legs meet, and an inner arm $L_{\mathrm{in}}$ to a terminal of the coupler junction. Each
qubit loop therefore shares one leg with the coupler loop, while
the two qubit loops share no branch. The qubits interact through the coupler
loop and through node 4, whose phase enters the inductive energy of both. The
circuit is symmetric under exchange of the two qubits.

Dropping the global phase leaves \vModesTwoQ{} normal modes
(Appendix~\ref{app:modes} lists $\omega_\mu$ and $\Omega_\mu$ for each). By
this symmetry, two of them, the qubit mode pair, are the symmetric and antisymmetric combinations of
the two qubit-junction phases. Their $\omega_\mu$ lie near \vDoubletFreqTwoQ,
close to the plasma frequency of one fluxonium linearized at zero phase and far
above a qubit's transition at half a flux quantum.

The coupler flux $\Phi_c$ tilts both qubits away from half a flux quantum by
the same amount, through the leg each qubit loop shares with the coupler loop,
which raises the gap $E_1-E_0$ between the lowest two levels of the coupled
circuit to as much as \vFZeroOneTwoQZThirty{} on the bias grid of
Sec.~\ref{sec:q2}. It also tunes the interaction between the qubits and with
it their static $ZZ$ shift~\cite{mundada_suppression_2019,
moskalenko_high_2022, ding_high-fidelity_2023}. At $\Phi_c = 0$ and
$\Phi_0/2$ the tilt vanishes and the interaction exceeds the qubits' own tunnel
splitting, so the lowest two levels are nearly degenerate.

\subsection{Three fluxonia with two tunable couplers}

The three-fluxonium circuit, Fig.~\ref{fig:circuits}(b), is built by the same
rule, with one flux-tunable coupler between each neighboring pair of qubits.
Every qubit loop has the same inductance (Appendix~\ref{app:params}), and an
isolated fluxonium with these values has $f_{01} = {}$\vFZeroOneThreeQ{} at
half a flux quantum.

Dropping the global phase leaves \vModesThreeQ{} normal modes with $\omega_\mu$
from \vModeSpanThreeQ{} (Appendix~\ref{app:modes}): five low modes that carry most of
the qubit-junction weight, two intermediate modes and the four highest, which we call stiff. The lower stiff
pair carries the two coupler junctions. The intermediate pair and the
upper stiff pair are nearly harmonic, the junctions holding only a few percent of their inductive energy.
In this basis no junction sits on one mode, and the participation ratios run
from \vEffModesThreeQ{} modes (Appendix~\ref{app:modes}).

\section{Two fluxonia: validation against exact diagonalization}\label{sec:q2}

We verify the tensor-network method against an exact solve of the same
truncated space and measure what both solvers cost at this size. Three checks
on the two-fluxonium circuit of Sec.~\ref{sec:circuits} come first. First, a
second implementation of the normal-mode Hamiltonian, written within
our group with the same conventions, reproduces the lowest
\vImplParityLevels{} levels of ours to \vImplParityTwoQ{} at
\vDimDDenseAnchorLow{} Fock levels per mode, along two sweeps of the coupler
flux, one of them with the qubit loops off half a flux quantum. Second, the
MPO, contracted to a dense matrix, has the eigenvalues of the
directly assembled Hamiltonian to \vMpoVsDenseTwoQSweeps{} along both sweeps.

\begin{figure}[t]
\centering
\includegraphics[width=\linewidth]{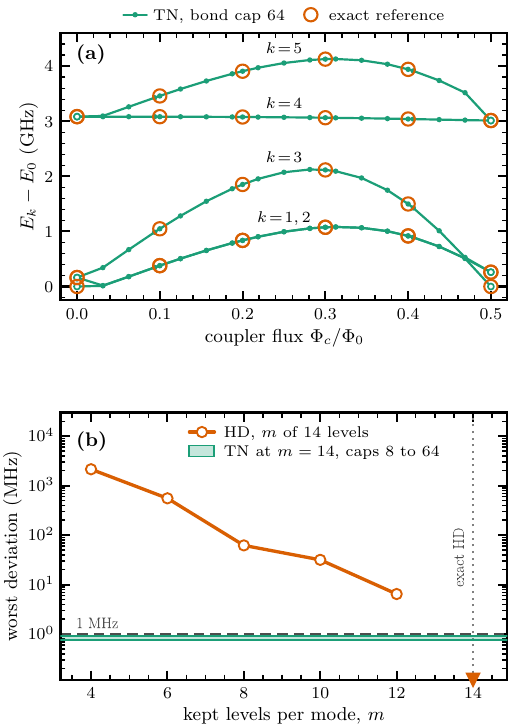}
\caption{(a) The lowest six levels of the two-fluxonium circuit against the
coupler flux, both qubit loops at $\Phi_0/2$, at the working cutoff of
\vDimDCertTwoQ{} levels per mode: dots are TN solves at bond cap 64 joined by
lines, rings an exact solve of the same space at the \vGridPointsTwoQ{} grid
points, and $k$ counts levels in energy order. The TN dots are open at
$\Phi_c = 0$ and $\Phi_0/2$, where the two lowest levels are nearly degenerate
and our DMRG schedule does not resolve their splitting (see text).
(b) Worst transition deviation among the lowest six levels over the grid, from
the reference, an exact solve of the same truncated space, so its Fock error is
not included. Circles: HD in the bare normal-mode basis, keeping $m$ levels per
mode. They trace the Fock convergence of that basis, which a TN at the same $m$
shares. No uniform cutoff $m=4$ to 12 enters the \vTolMHz{} class (dashed line), and
exact HD (dotted line; triangle at the axis floor) lies \vHdErrFourteen{}
from the reference. Band: the TN at $m=\vDimDCertTwoQ$ and bond caps 8 to 64,
inside the class between \vTnVsExactHdFourteen{} (cap 64) and \vTnErrSixteen{}
(cap 16).}
\label{fig:tn-vs-hd}
\end{figure}

Third, we test the solver on the \vGridPointsTwoQ{} grid points of
Fig.~\ref{fig:tn-vs-hd}(a), numbered 1 to 6 in increasing $\Phi_c$, both qubit
loops at $\Phi_0/2$; by symmetry they span a full period of the coupler flux. At
\vDimDDenseAnchorHigh{} Fock levels per mode, a space small enough for dense
diagonalization, the tensor network (TN) at bond cap 64 matches it at every
grid point, to
\vTnVsDenseEd{} on the transitions among the lowest six levels and
\vTnVsDenseEdEZero{} on the ground-state energy. At the working cutoff of
\vDimDCertTwoQ{} levels per mode both implementations ran a matrix-free exact
solve at every grid point; they agree to \vHdErrFourteen, and the second
implementation's solve is the reference for every deviation below.

The baseline is hierarchical diagonalization (HD) in the same bare normal-mode
basis as the TN. With each mode as one subsystem keeping $m$ Fock levels, no
partition has to be chosen and only the solver differs. A harmonic subsystem's
eigenstates are its Fock states, so HD at every cutoff $m$ of its ladder
(Sec.~\ref{sec:intro}), the uniform truncation $[m]^{\vModesTwoQ}$, is exact
diagonalization of its truncated space. At the working cutoff
$m=\vDimDCertTwoQ$ it is our own matrix-free Lanczos
solve~\cite{lanczos_iteration_1950} (Appendix~\ref{app:hardware}), which we
call exact HD. This cutoff misses the Fock plateau criterion of
Appendix~\ref{app:convergence} at \vTwoQFockMissCount{} of the
\vGridPointsTwoQ{} grid points and is not resolved at $\Phi_c = 0$ and
$\Phi_0/2$. Raising it to \vFockFloorTopDim{} levels per
mode moves the worst transition of the TN at cap 64 by \vFockFloorBand{} across
the grid. Every deviation below is measured against the reference at the working
cutoff and so leaves this Fock error out.

Figure~\ref{fig:tn-vs-hd} shows the lowest six levels along the coupler flux at
the working cutoff (a) and compares the solvers in that space of about
\vHilbertTwoQFourteenApprox{} states (b). The TN stays inside the \vTolMHz{}
class at every grid point from bond cap 8 to 64; at cap 64 its worst transition
deviation is \vTnVsExactHdFourteen{} (\vTnVsExactHdFourteenGround{} on the
transitions from the ground state), and its ground-state energy lies at most
\vTnEZeroOffsetFourteen{} above that of the reference. The bond dimension $\chi$
of Sec.~\ref{sec:truncation} is \vChiTwoQ{} here, below the smallest cap we
timed. The largest deviations fall at $\Phi_c = 0$ and $\Phi_0/2$, where the
lowest two levels are nearly degenerate (Sec.~\ref{sec:circuits}) and our DMRG
schedule does not resolve their splitting; at the other four points no cap
deviates by more than \vTnResidualOtherPoints. At $\Phi_c = 0$, a rerun with
\vTnProbeSweeps{} sweeps instead of \vSweepsTwoQ{} and a local Krylov dimension
of \vTnProbeKrylovDim{} instead of \vKrylovDim{} agrees with the reference to
\vTnProbeResidual{} at cap 64; it was not timed, and every cost below uses the
shorter schedule.

Every cost was measured as an isolated process on one CPU node
(Appendix~\ref{app:hardware}), with the thread
counts of each solver tuned on that host. Wall times are sums over grid points 2
to 6, since point 1 also carries the compilation (all six in
Table~\ref{tab:cost}); peak memory, the most RAM the process held, is net of
the process baseline, at one grid point.

\begin{figure*}[t]
\centering
\includegraphics[width=\textwidth]{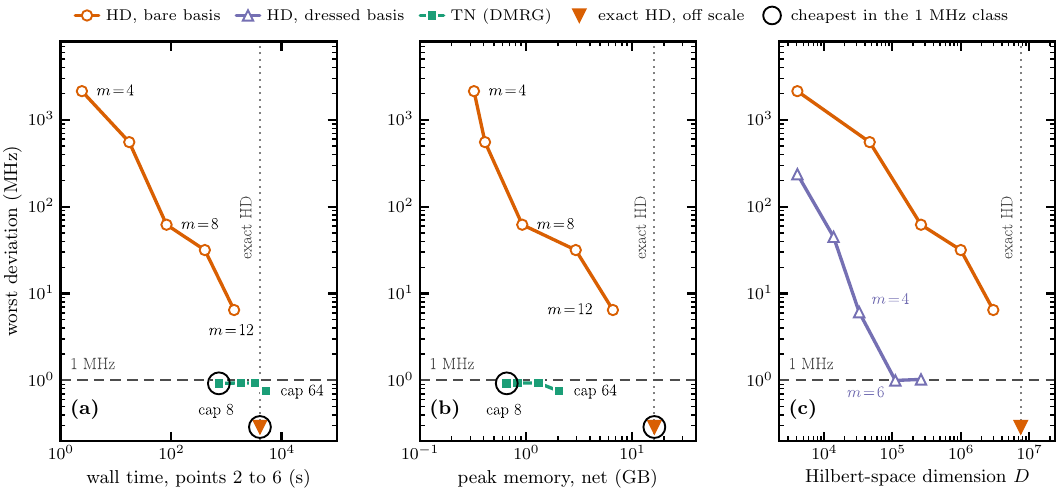}
\caption{The cost of reaching the \vTolMHz{} class on the two-fluxonium circuit
at the working cutoff, on the CPU node of Appendix~\ref{app:hardware}. Wall
time: sum over grid points
2 to 6; peak memory: one grid point, net of the process baseline. Deviation,
reference and dashed line as in Fig.~\ref{fig:tn-vs-hd}(b). Circles: bare-basis
HD at $m$ levels per mode; dotted vertical line and filled triangle: exact
HD, drawn at the axis floor; squares: the TN at bond caps 8 to 64; rings, in
(a) and (b): the cheapest configuration of each solver in the class. (a) Against
wall time: caps 8 to 32 take less than exact HD, cap 64 more. (b) Against
peak memory: every TN cap stays below the \vHdExactMem{} of exact HD,
which reflects the settings of our Krylov solver. (c) Against the Hilbert-space
dimension $D$: bare-basis HD reaches the class only at the working cutoff, while HD
in the dressed basis of the second implementation (open triangles: $m$ levels
on each of its three dressed junction subsystems, \vDressedHubDim{} on each
other mode) approaches the reference at smaller $D$. Its deviation measures
agreement between two differently truncated Hamiltonians, not accuracy in the
bare space, so it carries no ring, and the panel compares no cost.}
\label{fig:cost-scaling}
\end{figure*}

With \vModesTwoQ{} modes HD has the cheaper study (Table~\ref{tab:cost}): its
ladder from $m=4$ to the working cutoff, with no higher cutoff and no replicas,
takes \vLadderRatio{} of the time of the timed TN study. Run for run exact HD
also beats the TN at cap 64, at \vSpeedupProdWall{} of its wall time but
\vSpeedupProdMem{} times its memory. At caps 8 to 32 a single TN run is cheaper
(Fig.~\ref{fig:cost-scaling}): against the TN at cap 8, the smallest timed cap that
keeps every grid point in the class, exact HD takes \vSpeedupBareWall{}
times the wall time and \vSpeedupBareMem{} times the memory. The
\vHdExactMem{} of exact HD reflects the settings of our solver (Appendix~\ref{app:hardware}), whose tolerance lies far
below the class; its cost at a matched tolerance was not measured. These factors
price the bare basis, in which no uniform cutoff below the working one reaches
the class, as Fig.~\ref{fig:tn-vs-hd}(b) shows.

Seeding each grid point with the states converged at the previous one, a
standard continuation~\cite{hu_excited-state_2015}, saved little here. We ran
one chain per bond cap along the grid, with checks fixed before the runs; most
seeded steps failed a check and were re-solved cold. With every attempt
counted, the chains took less wall time than the cold solves above at caps 8 to
32 and more at cap 64, on the same host and threads. The seeded solves that
passed lie within \vWarmColdDevMax{} of the cold ones but farther from the
reference (Appendix~\ref{app:seeding}).

\begin{table}[t]
\centering
\caption{Cost of the two-fluxonium solves at the working cutoff, on the CPU
node of Appendix~\ref{app:hardware}: wall time over grid points 2 to 6 (all six in
parentheses), peak memory net of the process baseline; study totals over all
six. HD is in the bare normal-mode basis, where each run is an exact solve at
$m$ levels per mode. The TN Fock ladder runs four cutoffs up to \vFockFloorTopDim{} levels per
mode; the TN study leaves out the finer bond ladder for $\chi$, a second
replica set, the untimed Fock ladder and the reference solves.}
\label{tab:cost}
\begin{ruledtabular}
\begin{tabular}{lrr}
Solver & Wall time & Peak memory \\
\colrule
\multicolumn{3}{l}{\textit{Per run}} \\
TN, cap 8 & \vTnWallEight{} (\vTnWallEightSixPoint) & \vTnMemEight{} \\
TN, cap 64 & \vTnWallSixtyFour{} (\vTnWallSixtyFourSixPoint) & \vTnMemSixtyFour{} \\
Exact HD & \vHdWallExact{} (\vHdWallExactSixPoint) & \vHdExactMem{} \\
\colrule
\multicolumn{3}{l}{\textit{Study}} \\
HD, $m=4$ to \vDimDCertTwoQ{} & \vHdLadderTotal{} & \\
TN, caps 8 to 64 & \vTnLadderCaps{} & \\
TN, Fock ladder, cap 64 & \vTnLadderFock{} & \\
TN, \vReplicaCount{} replicas, cap 64 & \vTnLadderReplicas{} & \\
TN, total & \vTnLadderTotal{} & \\
\end{tabular}
\end{ruledtabular}
\end{table}

The dressed basis of the second implementation takes the phase across each
junction as one coordinate, in place of the normal modes, and builds that
coordinate's basis from its own oscillator and junction cosine. HD in that
basis approaches the reference with fewer states, as
Fig.~\ref{fig:cost-scaling}(c) shows. It truncates the
Hamiltonian differently, though, so its distance to the reference measures how
well two bases agree. The TN's distance is a solver error inside one truncated
operator.

\section{Three fluxonia at eleven normal modes}\label{sec:q3}

\subsection{The two truncations at the working point}
\label{sec:q3-convergence}

In the normal-mode basis of the three-fluxonium circuit of Sec.~\ref{sec:circuits} and
Fig.~\ref{fig:circuits}(b), no junction sits on one mode. The three qubit junctions
spread over the five low modes, and the worst places only
\vTopModeWeightWorstThreeQ{} of its normalized weight on its strongest mode, as
Appendix~\ref{app:modes} shows. A dressed basis built on these modes would
first have to assign each junction to one mode, and here that assignment has no
natural answer. The dressed basis of Sec.~\ref{sec:q2}, with one coordinate per
junction, needs no such assignment. It was not run on this circuit, and this
work quotes it at two fluxonia only.
Every number below at the working cutoff is taken at one bias point, the
working point, with both couplers unbiased and every qubit at half a flux
quantum. There, every junction phase enters with offset 0 or $\pi$, so the
spectrum is even in each loop flux. Each coupler has a loaded screening parameter
just below 1 (Sec.~\ref{sec:circuits}). On the flux scans of
Appendix~\ref{app:params} it sits in a single well, which flattens at half a
flux quantum. The quoted
levels are the lowest eight, with the lowest four beside them. They are named
by count: that they are the eight states of three qubits is not established,
since a second excited level of one qubit may lie among them. Their smallest
spacing, \vThreeQMinLevelSpacing{}, exceeds every truncation shift measured at
the working point, so each level is matched by its index from one truncation
to the next. The working cutoff keeps \vDimDLowThreeQ{} Fock levels on each
of the five low modes of Sec.~\ref{sec:circuits} and \vDimDHighThreeQ{} on each
of the other six, a product space of \vHilbertThreeQProdApprox{} states. Exact
diagonalization of that space is out of reach as a method, in dense or sparse
storage (Sec.~\ref{sec:q3-cost}), and a matrix-free exact solve was not
attempted at this size, so this work has no exact reference at the working cutoff and characterizes
the eleven-mode solve on its own two truncation axes. On this circuit the MPO
construction, validated against exact diagonalization in Sec.~\ref{sec:q2},
reproduces a dense solve at \vDimDAnchorThreeQ{} Fock levels per mode to
numerical precision.

\begin{figure*}[t]
\centering
\includegraphics[width=\textwidth]{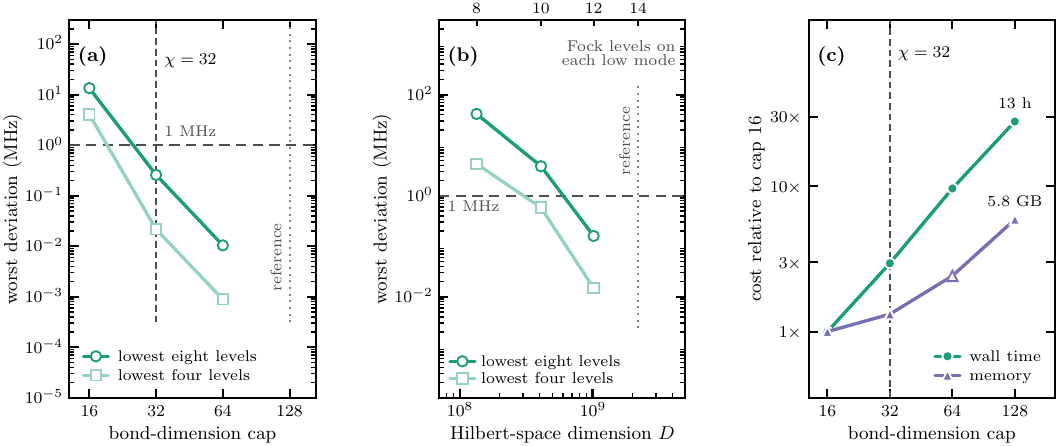}
\caption{The tensor-network solve of the three-fluxonium circuit at its working
point, both couplers unbiased and every qubit at half a flux quantum. (a) Bond
axis at the working cutoff: the worst transition deviation over all pairs of
the lowest eight levels (circles) and of the lowest four (squares) from the
solve at bond cap 128 (dotted line). Every step up the ladder from cap 32 on
moves the transitions by less than \vTolMHz{} (horizontal dashed line), so
$\chi = \vChiThreeQ$ (vertical dashed line). (b) Fock axis at bond cap 64: the five low modes are raised step by step
to \vDimDLowThreeQ{} levels (top axis), the other six kept at
\vDimDHighThreeQ{}, each step against \vDimDRefThreeQ{} levels (dotted line);
the same ladder at cap 32 agrees with this one to within
\vThreeQBondMoveMax{} (Appendix~\ref{app:convergence}). At the working cutoff the deviation is at most
\vFockErrAThree{} on the lowest eight levels and \vFockErrAThreeLowFour{} on
the lowest four, whether the bond cap is 32 or 64. (c) Wall time and peak memory, net of the process baseline, of
one solve per cap at the working cutoff, on the CPU node of Appendix~\ref{app:hardware},
each relative to its value at cap 16 (\vThreeQWallSixteen{} and \vThreeQMemSixteen{}), the cap-128
values labeled. The cap-64 memory (open triangle) is the high-water mark of a
process that solved several configurations, an upper bound. Over three
doublings the time grows by a factor of \vThreeQWallGrowth{} and the memory by
\vThreeQMemGrowth{}. At $\chi = \vChiThreeQ$ the solve takes
\vThreeQWallThirtyTwo{} and \vThreeQMemThirtyTwo{}, and the lowest eight levels
meet every criterion of Appendix~\ref{app:convergence} at one setting.}
\label{fig:q3}
\end{figure*}

Convergence is checked as for any truncated solve: each truncation is raised
in turn while the other is held, and the quoted transitions are compared with
those at the largest setting computed on that axis
(Sec.~\ref{sec:truncation}). In Fig.~\ref{fig:q3}(a) the bond axis raises the cap from 16 to 128 at
the working cutoff; in Fig.~\ref{fig:q3}(b) the Fock axis raises the five low
modes step by step to \vDimDLowThreeQ{} levels, each read against
\vDimDRefThreeQ{} levels, the other six held at \vDimDHighThreeQ{}. At the working cutoff both
distances lie well inside the \vTolMHz{} tolerance
(Table~\ref{tab:q3-convergence}).

\begin{table}[t]
\centering
\caption{Convergence checks of the three-fluxonium solve at the working point
and the working cutoff. Each entry except the replica spread is the largest
change of a transition between any two of the quoted levels, against the
largest setting of its axis
(Sec.~\ref{sec:truncation}); Fock-axis and probe entries are the worst over
bond caps 32 and 64. Every entry lies below the \vTolMHz{} tolerance of
Appendix~\ref{app:convergence}.}
\label{tab:q3-convergence}
\begin{ruledtabular}
\begin{tabular}{lrr}
Check & Lowest eight & Lowest four \\
\colrule
\multicolumn{3}{l}{\textit{Bond axis, against cap 128}} \\
Cap 32 & \vBondErrAThirtyTwo{} & \vBondErrLowFourThirtyTwo{} \\
Cap 64 & \vBondMoveThreeQ{} & \vBondMoveThreeQLowFour{} \\
\colrule
\multicolumn{3}{l}{\textit{Fock axis, against \vDimDRefThreeQ{} levels on the low modes}} \\
\vDimDLowThreeQ{} levels & \vFockErrAThree{} & \vFockErrAThreeLowFour{} \\
\colrule
\multicolumn{3}{l}{\textit{Off-path probes, one step from the working cutoff}} \\
Intermediate at \vDimDMidProbeThreeQ{} levels & \vFockOffPathMid{} & \vFockOffPathMidLowFour{} \\
Stiff at \vDimDStiffProbeThreeQ{} levels & \vFockOffPathStiff{} & \vFockOffPathStiffLowFour{} \\
\colrule
\multicolumn{3}{l}{\textit{Random starts at cap 32}} \\
Spread of \vReplicaCount{} replicas &\vReplicaSpreadThreeQ{} & \vReplicaSpreadThreeQLowFour{} \\
\end{tabular}
\end{ruledtabular}
\end{table}

Without an exact reference this check has weak points. A shrinking step
does not bound the error that remains. The variational solve can settle in a
local minimum. Each axis is measured at a finite setting of the other, so the
error of one truncation can be mistaken for that of the other. The Fock
ladder follows one path through
the per-mode cutoffs.

The bond ladder converges fast over the caps run: each
doubling of the cap moves the transitions less than the one before, by a
factor of \vThreeQBondStepFallFirst{} and then
\vThreeQBondStepFallSecond{}. At cap 32, \vReplicaCount{} solves from random states with distinct
seeds agree to \vReplicaSpreadThreeQ{}. The Fock ladder ran at caps 32 and 64,
and between them the Fock error at every cutoff moves by at most
\vThreeQBondMoveMax{}, so over these two caps the Fock error does not
depend on the bond cap. Off that path
the two intermediate and the four stiff modes of Sec.~\ref{sec:circuits} were
each raised by one step; the stiff modes carry the largest shift, and one step
cannot show a plateau. The worst occupancy of the highest kept Fock level,
\vTopFockThreeQ{}, lies below the \vTopFockCrit{} flag; on the stiff modes it
is no reliable marker, since the stiff probe's own occupancy,
\vTopFockStiffProbe{}, exceeds it (Appendix~\ref{app:convergence}).

The lowest eight levels thus meet every criterion of
Appendix~\ref{app:convergence} at one setting, the working cutoff at bond
cap \vChiThreeQ{}. The lowest four meet them too, by a wider margin on the
Fock, bond and replica checks. Appendix~\ref{app:convergence} lists the
transitions of the lowest eight levels with the move along each axis. The quoted transitions are those of the cap-128 solve, which
the working solve at cap \vChiThreeQ{} matches to within \vBondErrAThirtyTwo{};
the Fock error measured at caps 32 and 64 is carried to cap 128, an assumption
Appendix~\ref{app:convergence} states with its evidence.

\subsection{Memory and run time}
\label{sec:q3-cost}

Exact diagonalization in the same basis is out of reach as a method at eleven
modes. A dense complex matrix of the truncated space at the working cutoff
would occupy \vEdDenseMemThreeQ{}. Sparse storage would still exceed the
memory of the CPU node. In a charge basis a junction cosine only shifts the
charges of the two nodes it joins, adding a few entries per row. In the
normal-mode basis every junction acts on all eleven modes
(Appendix~\ref{app:modes}), and its cosine, Eq.~\eqref{eq:junction-product},
is a product of single-mode matrices on all of them. The inductive matrix
$E_L$ also couples every pair of modes, so each row holds at least one entry
per pair of modes before any junction term is added. A matrix-free solver stores no matrix, only vectors
of the space, and so needs far less memory than either form of storage. No
matrix-free exact solve was attempted at this size, though: ours, at the
settings of Appendix~\ref{app:hardware}, would need about \vHdElevenModeMem{} there
(Table~\ref{tab:q3-cost}), above the CPU node's \vNodeMem{} of RAM. In the Fock
basis of the normal modes the Hamiltonian is real symmetric at every flux bias,
as checked numerically at two fluxonia, so its stored vectors could be real, at half the memory of complex
ones; lower-memory implementations of this kind are under development and
testing.

Each wall time and memory in this paragraph was measured per solve, at the working
point, on the CPU node of Appendix~\ref{app:hardware}. Memory is net of the
process baseline, as in Table~\ref{tab:q3-cost}. Establishing the truncation is
a one-time study. Its bond
ladder at the working cutoff took \vThreeQLadderHours{} over four solves. Its
Fock axis took \vThreeQFockSweepSolves{} solves: four cutoffs of the low modes
and the two probes, each at bond caps 32 and 64. The largest of them, the stiff
probe's, spans about \vHilbertThreeQStiffProbe{} states. In this study, wall
time grows much faster than memory: over three doublings of the cap the time
grows by a factor of \vThreeQWallGrowth{} and the memory by
\vThreeQMemGrowth{}, as Fig.~\ref{fig:q3}(c) shows. No solve of it needed more than
\vThreeQMemOneTwentyEight{}, and the Fock axis at most
\vThreeQFockSweepMaxMem{}. Each bias point then needs one solve at the
working setting, which takes
\vThreeQWallThirtyTwo{} and \vThreeQMemThirtyTwo{} of memory on the CPU node. A
flux sweep of ten bias points therefore runs overnight. This work does not establish that the setting found at the working
point also holds at the other bias points of such a sweep.

\begin{table}[t]
\centering
\caption{Cost of the three-fluxonium solves at the working cutoff, on the
CPU node of Appendix~\ref{app:hardware}: wall time of one solve at the working point, peak memory
net of the process baseline; the cap-64 memory is an upper bound, its process
having solved other configurations. The study total sums the four solves of
the bond ladder; the Fock-axis and replica solves are not summed, since some
ran on other hosts. The HD row is not a run but bare-basis HD carried from six modes
(Fig.~\ref{fig:memory}), the one extrapolation in this work.}
\label{tab:q3-cost}
\begin{ruledtabular}
\begin{tabular}{lrr}
Solver & Wall time & Peak memory \\
\colrule
\multicolumn{3}{l}{\textit{Per run}} \\
TN, cap 16 & \vThreeQWallSixteen{} & \vThreeQMemSixteen{} \\
TN, cap 32 & \vThreeQWallThirtyTwo{} & \vThreeQMemThirtyTwo{} \\
TN, cap 64 & \vThreeQWallSixtyFour{} & \vThreeQMemSixtyFour{} \\
TN, cap 128 & \vThreeQWallOneTwentyEight{} & \vThreeQMemOneTwentyEight{} \\
HD, not run & \vHdGuideHours{} & \vHdElevenModeMem{} \\
\colrule
\multicolumn{3}{l}{\textit{Study}} \\
TN, caps 16 to 128 & \vThreeQLadderHours{} & \\
\end{tabular}
\end{ruledtabular}
\end{table}

\begin{figure}[!tbp]
\centering
\includegraphics[width=\linewidth]{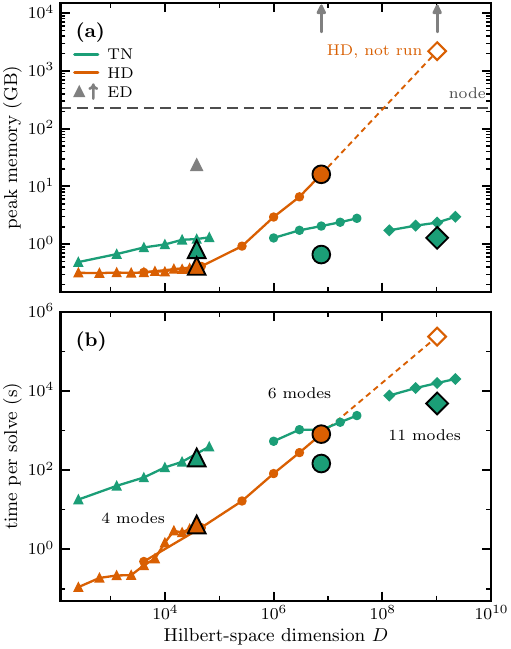}
\caption{Memory and run time against Hilbert-space dimension $D$ for two
fluxonia at \vModesFourMode{} modes (triangles, Appendix~\ref{app:params}) and \vModesTwoQ{} modes (circles)
and three fluxonia at \vModesThreeQ{} modes (diamonds): (a) peak memory net of
the process baseline, (b) wall time per solve at one bias point. Lines join one
solver on one circuit: the TN (green) at bond cap 64 along the Fock cutoff, and
HD in the bare basis (orange) along its kept levels per mode up to the exact
solve. Large outlined markers: each circuit's cheapest run in the \vTolMHz{}
class, against an exact solve at \vDimDCertTwoQ{} levels per mode at four and
six modes and against the cap-128 and \vDimDRefThreeQ{}-level solves at eleven. In (a), exact diagonalization (ED, gray),
which stores a dense complex matrix, is on the scale only at four modes, and
arrows mark the others. Dashed line: the CPU node's \vNodeMem{} of RAM. The open
orange diamonds and dashed orange lines are not runs but HD carried to eleven
modes, the one extrapolation in this work (Sec.~\ref{sec:q3-cost}). Four-mode
times exclude the Julia compilation, \vCompileTnFourMode{} for the TN and
\vCompileHdFourMode{} for HD.}
\label{fig:memory}
\end{figure}

Figure~\ref{fig:memory} places both circuits, with a smaller two-fluxonium
circuit of four modes (Appendix~\ref{app:params}), on one axis of Hilbert-space dimension, with the
bare-basis HD solver of Sec.~\ref{sec:q2} at eleven modes only as implemented,
not run. From six modes on, the memory and run time of the TN grow far more
slowly with $D$ than those of HD, which is the cheaper solver at four modes.
Exact HD needs \vHdBytesPerState{} per basis state, a figure set
by the storage settings of our solver. Carried to eleven modes at that
footprint it would need \vHdElevenModeMem{}, and the six-mode HD times at $m = 8$
to $14$, carried to eleven modes along their trend in $D$, reach
\vHdGuideHours{} per solve there. At eleven modes the outlined marker is the
working solve, identified against the cap-128 and \vDimDRefThreeQ{}-level solves of
Sec.~\ref{sec:q3-convergence}; at four modes it is set against
\vDimDFourMode{} levels per mode, whose Fock truncation is uncertain by about
\vFourModeSixteenVsFourteen{} on the lowest six levels. The two-fluxonium and
three-fluxonium solves ran with different thread settings and sweep schedules
(\vSweepsTwoQ{} and \vSweeps{} sweeps), so they are not compared wall for wall.

The bond dimension
$\chi$ of Sec.~\ref{sec:truncation} is \vChiTwoQ{} on the lowest six
levels of the two-fluxonium circuit and \vChiTwoQLowFour{} on its lowest
four, per point of the grid of Sec.~\ref{sec:q2}, and \vChiThreeQ{} on the
lowest eight levels of the three-fluxonium circuit and \vChiThreeQLowFour{} on
its lowest four, at one bias point. At that point cap 16 misses, so the
smallest passing cap lies between 16 and 32. For scale, a matrix product state that can
represent any state of the three-fluxonium space at the working cutoff needs,
at each cut of the chain, a bond dimension equal to the dimension of the
smaller side~\cite{schollwoeck_density-matrix_2011}. In the site order of
Appendix~\ref{app:convergence} the largest such value is
\vExactMpsBondThreeQ{}, between the four lowest modes and the other seven. At
cap \vChiThreeQ{} a state is stored in at most \vMpsNumbersThreeQ{} numbers,
against \vHilbertThreeQProdApprox{} for a dense vector. At the same absolute tolerance the larger
circuit needs a larger bond cap, by a factor of \vChiGrowthRange{}. This
factor compares two circuits that differ in size, layout and coupler regime
(Sec.~\ref{sec:circuits}). They were solved with \vSweepsTwoQ{} and \vSweeps{}
sweeps and a fixed site ordering. Optimizing the ordering can change the bond
dimension DMRG needs~\cite{legeza_optimizing_2003}.

\section{Discussion}\label{sec:disc}

In the bare normal-mode basis, hierarchical diagonalization truncates only the number of Fock levels kept per mode. At two fluxonia no uniform cutoff below the working one comes within \vTolMHz{} of the exact solve, as Fig.~\ref{fig:tn-vs-hd}(b) shows. The tensor network makes the same basis affordable. It keeps each mode's truncated Fock space in full and truncates only the bond dimension, so the cost of a solve is set by the number of modes, the bond dimension and the per-site dimension (Fig.~\ref{fig:memory}). At two fluxonia the tensor network agrees with exact diagonalization of the same truncated space to within \vTolMHz{} on the lowest six levels (Sec.~\ref{sec:q2}). At three fluxonia, with the same method and implementation, the lowest eight levels at one bias point meet every criterion of Appendix~\ref{app:convergence} at one setting (Sec.~\ref{sec:q3-convergence}), and one solve at that setting's bond dimension, \vChiThreeQ, takes \vThreeQWallThirtyTwo{} and \vThreeQMemThirtyTwo{}. Each three-fluxonium result is checked by the two truncation sweeps, each read against its largest setting, and both sweeps remain affordable at eleven modes, on one CPU node (Sec.~\ref{sec:q3-cost}). There, with \vHilbertThreeQProdApprox{} states, exact diagonalization in the same basis is out of reach as a method, in dense or sparse storage, and a matrix-free exact solve was not attempted at that size. At bond dimension \vChiThreeQ{} the tensor network stores each state of that space in a small fraction of the numbers a dense vector needs (Sec.~\ref{sec:q3-cost}).

From the two-fluxonium circuit of Sec.~\ref{sec:q2} to the three-fluxonium circuit, the bond cap needed at the same absolute tolerance grows by a factor of \vChiGrowthRange{} (Sec.~\ref{sec:q3-cost}), between circuits that differ in size, layout and coupler regime. Why a bond dimension this small suffices, although every junction of the three-fluxonium circuit acts on several modes at once, is left open. The \vTolMHz{} class is a tolerance on each transition, so it does not settle an interaction between the qubits at or below its scale, such as a residual static $ZZ$ shift (Sec.~\ref{sec:circuits}).

Reducing the computed spectrum to an effective qubit Hamiltonian without a perturbative expansion in the coupling~\cite{schrieffer_relation_1966, bravyi_schrieffer-wolff_2011, consani_effective_2020} is deferred to separate work. On the solver side, a local basis smaller than the bare Fock basis, found by an algorithm from the circuit's inputs, would lower the cost while keeping the method generic. A dressed basis built on the normal modes is the obvious starting point, and the junction-to-mode assignment it needs is the step such an algorithm would have to supply. DMRG could also target all quoted levels at once~\cite{baker_direct_2024, gonzalez-garcia_multi-target_2025}, and the mode ordering could be chosen from quantum-information entropies~\cite{legeza_optimizing_2003}. A reimplementation of the tensor-network solver aimed at shorter turnaround times is under development and testing and will be reported separately.

\begin{acknowledgments}
We thank Tim Duty, Andrea Pérez, Victor Sanchez Gimenez, David López Núñez,
Henry Chew, Gonzalo Troncoso, Volkan Tozan and Sergi Terradas-Briansó of the
Qilimanjaro design team for the circuit designs and for their support. We also
thank Jordi Riu and Sofyan Iblisdir for fruitful discussions.

The authors acknowledge RES resources provided by the Barcelona Supercomputing
Center on MareNostrum~5 under allocation INNO-2026-1-0004. Benchmark runs were
performed on in-house CPU nodes.

We used the AI assistant Claude Code (Anthropic) to develop the simulation
and plotting code and to assist with the manuscript text. The authors
reviewed all AI output and take full responsibility for the content of this
article.
\end{acknowledgments}

\section*{Data Availability}
Data are available from the authors upon reasonable request.

\appendix

\section{Circuit parameters}
\label{app:params}

\begin{figure*}[tbp]
\centering
\makebox[\textwidth][l]{\textbf{(a)}}\\[1mm]
\includegraphics[width=0.62\textwidth]{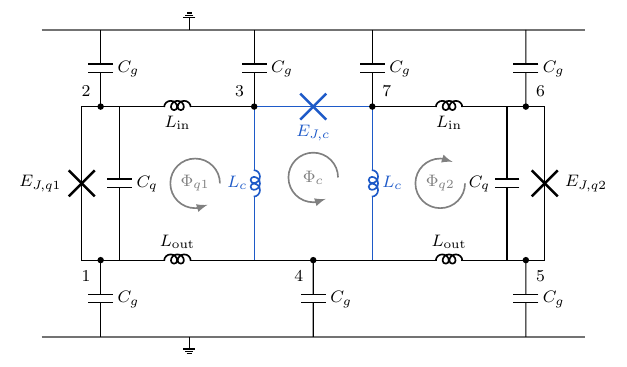}\\[4mm]
\makebox[\textwidth][l]{\textbf{(b)}}\\[1mm]
\includegraphics[width=0.96\textwidth]{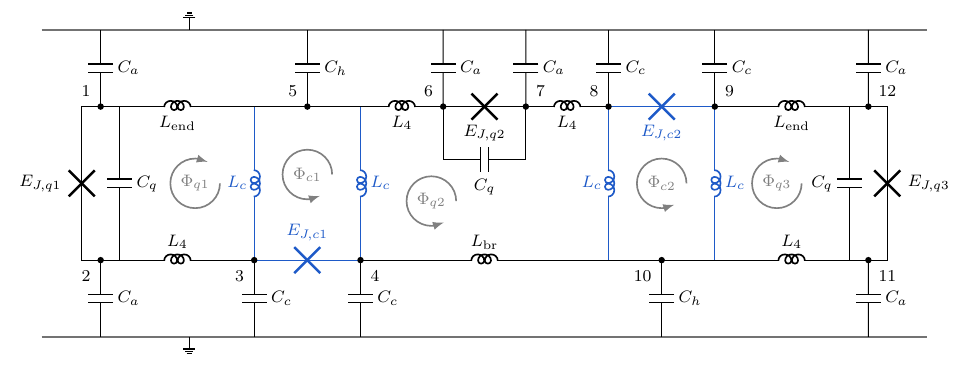}
\caption{(a) The two-fluxonium circuit and (b) the three-fluxonium circuit of
Fig.~\ref{fig:circuits}, with the two element classes omitted there restored:
the capacitance from every node to ground and the external-flux loop of every
junction. The ground capacitances are drawn to two ground rails that are the
same node, and the inductors alone span the circuit nodes, so each inductive
network is floating. Each junction closes exactly one independent flux loop,
drawn as a gray arc labeled with its external flux, and neighboring loops
share one inductive leg. The sign of each external flux $\Phi_\ell$ is fixed
by the order $i(b)$, $j(b)$ in which Fig.~\ref{fig:circuits} lists the
junction's branch. Element values are in Table~\ref{tab:params}.}
\label{fig:circuits-full}
\end{figure*}

Table~\ref{tab:params} lists every element of both benchmark circuits and of
the four-mode circuit described at the end of this appendix, and
Fig.~\ref{fig:circuits-full} draws the benchmark circuits in full. With the
node numbering of
Fig.~\ref{fig:circuits-full}, the table fixes the capacitance matrix $C$, the
geometric inverse-inductance matrix $L^{-1}$ and the junction
list, the entire input of Sec.~\ref{sec:method}, so every mode frequency of
Table~\ref{tab:modes} can be recomputed from this appendix alone.

\begin{table*}[tbp]
\centering
\caption{Elements of the two benchmark circuits and of the four-mode circuit.
Branches and nodes are given in the 1-based numbering of
Figs.~\ref{fig:circuits} and~\ref{fig:circuits-full} and, for the four-mode
circuit, of the closing paragraph of this appendix. Josephson energies are
$E_J/h$. Design inputs are printed as given. The four-mode inductances and
capacitances, the latter the node-to-ground and mutual entries of a simulated
Maxwell matrix, are rounded to three significant figures. The indented rows are derived from them: a loop inductance is
the sum of the inductances the loop traverses, and
$E_L = \varphi_0^2/(hL)$ with $\varphi_0^2/h = {}$\vFluxQuantumConst,
each computed from the unrounded inputs and rounded to three significant
figures.}
\label{tab:params}
\begin{ruledtabular}
\begin{tabular}{llr}
Element & Branches or nodes & Value \\
\colrule
\multicolumn{3}{l}{\textit{Two fluxonia, \vNodesTwoQ{} nodes, \vModesTwoQ{} normal modes}} \\
qubit junctions $E_{J,q1}$, $E_{J,q2}$ & $(1,2)$, $(5,6)$ & \vEjQubitTwoQ{} \\
coupler junction $E_{J,c}$ & $(3,7)$ & \vEjCouplerTwoQ{} \\
outer body arms $L_{\mathrm{out}}$ & $(1,4)$, $(4,5)$ & \vLBodyTwoQ{} \\
inner body arms $L_{\mathrm{in}}$ & $(2,3)$, $(6,7)$ & \vLBodyTwoQ{} \\
coupler legs $L_c$ & $(3,4)$, $(4,7)$ & \vLCouplerLegTwoQ{} \\
junction shunts $C_q$ & $(1,2)$, $(5,6)$ & \vCshunt{} \\
node-to-ground $C_g$ & every node & \vCgroundTwoQ{} \\
\quad qubit loop & $L_{\mathrm{out}}+L_{\mathrm{in}}+L_c$ & \vLoopQubitTwoQ{} \\
\quad qubit $E_L$, $E_J/E_L$ & & \vElQubitTwoQ, \vEjElQubitTwoQ{} \\
\quad qubit $E_C$ & & \vEcQubitTwoQ{} \\
\quad coupler loop & $2L_c$ & \vLoopCouplerTwoQ{} \\
\quad coupler $E_L$, $E_J/E_L$ & & \vElCouplerTwoQ, \vEjElCouplerTwoQ{} \\
\colrule
\multicolumn{3}{l}{\textit{Three fluxonia, \vNodesThreeQ{} nodes, \vModesThreeQ{} normal modes}} \\
qubit junctions $E_{J,q1..q3}$ & $(1,2)$, $(6,7)$, $(11,12)$ & \vEjQubitThreeQ{} \\
coupler junctions $E_{J,c1}$, $E_{J,c2}$ & $(4,3)$, $(8,9)$ & \vEjCouplerThreeQ{} \\
body quarter arms $L_4$ & $(2,3)$, $(5,6)$, $(7,8)$, $(10,11)$ & \vLFourThreeQ{} \\
coupler legs $L_c$ & $(3,5)$, $(4,5)$, $(8,10)$, $(9,10)$ & \vLCouplerLegThreeQ{} \\
end-around arms $L_{\mathrm{end}}$ & $(1,5)$, $(12,9)$ & \vLEndThreeQ{} \\
rail bridge $L_{\mathrm{br}}$ & $(4,10)$ & \vLBridgeThreeQ{} \\
junction shunts $C_q$ & $(1,2)$, $(6,7)$, $(11,12)$ & \vCshunt{} \\
qubit-terminal $C_a$ & $1,2,6,7,11,12$ & \vCqubitThreeQ{} \\
coupler-terminal $C_c$ & $3,4,8,9$ & \vCcouplerThreeQ{} \\
coupler-leg nodes $C_h$ & $5,10$ & \vChubThreeQ{} \\
\quad qubit loop, outer & $L_{\mathrm{end}}+L_c+L_4$ & \vLoopQubitThreeQ{} \\
\quad qubit loop, central & $2L_4+2L_c+L_{\mathrm{br}}$ & \vLoopQubitThreeQ{} \\
\quad qubit $E_L$, $E_J/E_L$ & & \vElQubitThreeQ, \vEjElQubitThreeQ{} \\
\quad coupler loop & $2L_c$ & \vLoopCouplerThreeQ{} \\
\quad coupler $E_L$, $E_J/E_L$ & & \vElCouplerThreeQ, \vEjElCouplerThreeQ{} \\
\quad qubit $E_C$; isolated qubit $f_{01}$ at $\Phi_q = \Phi_0/2$ & & \vEcQubitThreeQ, \vFZeroOneThreeQ{} \\
\colrule
\multicolumn{3}{l}{\textit{Two fluxonia with SQUIDs, \vNodesFourMode{} nodes, \vModesFourMode{} normal modes}} \\
qubit SQUIDs $E_{J,q1}$, $E_{J,q2}$, per arm & $(1,2)$, $(3,4)$ & \vEjQubitFourMode{} \\
coupler SQUID $E_{J,c}$, per arm & $(2,3)$ & \vEjCouplerFourMode{} \\
qubit arms $L_q$ & $(1,5)$, $(4,5)$ & \vLQubitFourMode{} \\
coupler legs $L_c$ & $(2,5)$, $(3,5)$ & \vLCouplerLegFourMode{} \\
node-to-ground $C_g$ & $1$ to $5$ & \vCgroundFourMode{} \\
to the center node & $(1,5)$ to $(4,5)$ & \vCcenterFourMode{} \\
between neighbors & $(1,2)$, $(2,3)$, $(3,4)$ & \vCneighborFourMode{} \\
between distant nodes & $(1,3)$, $(2,4)$, $(1,4)$ & \vCdistantFourMode{} \\
\quad qubit loop & $L_q+L_c$ & \vLoopQubitFourMode{} \\
\quad qubit $E_L$ & & \vElQubitFourMode{} \\
\quad coupler loop & $2L_c$ & \vLoopCouplerFourMode{} \\
\quad coupler $E_L$ & & \vElCouplerFourMode{} \\
\end{tabular}
\end{ruledtabular}
\end{table*}

The three qubit loops of the three-fluxonium circuit close on the same
inductance although they traverse different branches: the two outer qubits
run through an end-around arm, the central one through the rail bridge. That
is by construction, since $L_{\mathrm{end}} = 3L_4 + 2L_c$ and
$L_{\mathrm{br}} = 2L_4 + L_c$ exactly, so every qubit loop traverses
$4L_4 + 3L_c$ whichever branches it follows, and all three qubit loops have
the same $E_L$. The circuit is not mirror symmetric, so the two outer qubits are
not equivalent: the end-around arm joins the first qubit to node 5 but the
third to coupler terminal 9, and the central qubit's arms and the rail bridge
also land differently on the two sides (Table~\ref{tab:params}). The two outer
qubit junctions differ in participation, \vEffModesThreeQQubitOne{} modes for
the first against \vEffModesThreeQQubitThree{} for the third
(Appendix~\ref{app:modes}).

The ratio $E_J/E_L$ of a coupler in Table~\ref{tab:params} counts only its two
legs, the inductance its junction sees with every other junction open. In the
circuit every other junction adds its Josephson inductance at the classical
minimum, so at the minima of the working point the coupler junction sees a
smaller inductance, and its screening parameter is \vBetaLoadedMinThreeQ{} to
\vBetaLoadedMaxThreeQ{}. With both couplers at half a flux quantum the
screening parameter is \vBetaLoadedMinHalfFluxThreeQ{} to
\vBetaLoadedMaxHalfFluxThreeQ{}. The classical potential has \vMinimaThreeQ{}
minima at the working point, one for each configuration of the three qubit
wells. On scans of both couplers together and of the first coupler alone, from
zero to half a flux quantum with the qubits at half a flux quantum, the
potential never has more than \vMinimaMaxThreeQ{} minima, and no two of them
share a configuration of the qubit wells. A bistable coupler would add a
second minimum to one configuration, so no coupler is bistable on these scans.

The four-mode points of Fig.~\ref{fig:memory} come from a third circuit, a
galvanically coupled fluxonium pair of the kind of
Ref.~\cite{weiss_fast_2022}, with SQUIDs in place of its single junctions and
a floating center node in place of its ground. Nodes 1 to 4 run along the
chain qubit, coupler, qubit, and each connects through one inductor to the
center node 5, so dropping the global phase leaves \vModesFourMode{} normal
modes (Table~\ref{tab:params}). A SQUID enters Eq.~\eqref{eq:lumped} as two
junctions on the same nodes, with phase offsets
$2\pi(\Phi_\ell \mp \Phi_s/2)/\Phi_0$, where $\Phi_\ell$ threads the loop the
branch closes with the inductors and $\Phi_s$ the SQUID itself. The pair thus
acts as one junction of energy $2E_J\cos(\pi\Phi_s/\Phi_0)$. The capacitance
matrix is an electromagnetic simulation of a laboratory layout and includes
the junction capacitances. Every four-mode number is at one bias point: both
qubit loops at $\Phi_0/2$, both qubit SQUIDs at \vSquidFluxFourMode$\,\Phi_0$,
the coupler loop and SQUID unbiased. There the bond dimension $\chi$ of
Sec.~\ref{sec:truncation} is \vChiFourMode{} on the lowest six levels, and at
bond cap 64 the tensor network agrees with an exact solve of the space of
\vDimDFourMode{} levels per mode to \vTnVsExactFourMode. Raising the cutoff to
\vDimDFourModeTop{} levels per mode moves the lowest four levels by
\vFourModeSixteenVsFourteenLowFour{} and the lowest six by about
\vFourModeSixteenVsFourteen.

\section{Normal modes and junction participation}
\label{app:modes}

We list the normal-mode frequencies of both circuits, built in
Sec.~\ref{sec:mpo-modes} from the parameters of Appendix~\ref{app:params}, and
the participation of every junction in those modes, which gives the
participation ratio used in Sec.~\ref{sec:intro}.

\subsection{Two frequencies per mode}

Each mode carries two frequencies, and Table~\ref{tab:modes} lists both. The
first, $\omega_\mu$, is the square root of the generalized eigenvalue of
Eq.~\eqref{eq:normalmodes}, in which every junction is linearized about zero
phase and folded into $L^{-1}_{\mathrm{tot}}$; it belongs to the linear circuit
and does not depend on the bias. The second, $\Omega_\mu$ of
Sec.~\ref{sec:mpo-modes}, is the frequency of the oscillator whose Fock states
site $\mu$ is built on. It uses the geometric inductance alone, since the
junction enters through its full cosine, and it fixes the zero-point spread of
$\hat\phi_\mu$ and so the meaning of a Fock cutoff on that site. The two agree
where the junctions participate weakly and part where a junction is
inductively important, most of all on the qubit mode pair of the two-fluxonium
circuit: \vDoubletPencilPair{} as linearized modes, \vDoubletLocalPair{} as
local oscillators, a factor of \vDoubletConventionRatio, so a Fock cutoff
chosen from $\omega_\mu$ would refer to the wrong oscillator.

\begin{table*}[tbp]
\centering
\caption{Normal-mode frequencies of both circuits in ascending order of
$\omega_\mu$, to 1~MHz: the linearized $\omega_\mu$ of
Eq.~\eqref{eq:normalmodes} and the local oscillator $\Omega_\mu$ of
Sec.~\ref{sec:mpo-modes}, both computed from Table~\ref{tab:params} and
independent of the flux bias. The eleven modes of the three-fluxonium circuit
fall into five low, two intermediate and four stiff modes
(Sec.~\ref{sec:circuits}).}
\label{tab:modes}
\begin{ruledtabular}
\begin{tabular}{ll}
Circuit and convention & Frequencies (GHz) \\
\colrule
\multicolumn{2}{l}{\textit{Two fluxonia, \vModesTwoQ{} normal modes}} \\
\quad $\omega_\mu$, linearized  & \vModeFreqTwoQ \\
\quad $\Omega_\mu$, local oscillator & \vLocalOscTwoQ \\
\colrule
\multicolumn{2}{l}{\textit{Three fluxonia, \vModesThreeQ{} normal modes}} \\
\quad $\omega_\mu$, linearized, low modes & \vModeFreqThreeQLow \\
\qquad intermediate modes & \vModeFreqThreeQMid \\
\qquad stiff modes & \vModeFreqThreeQStiff \\
\quad $\Omega_\mu$, local oscillator, low modes & \vLocalOscThreeQLow \\
\qquad intermediate modes & \vLocalOscThreeQMid \\
\qquad stiff modes & \vLocalOscThreeQStiff \\
\end{tabular}
\end{ruledtabular}
\end{table*}

\subsection{Junctions across the modes}

\begin{figure*}[tbp]
\centering
\includegraphics[width=0.97\textwidth]{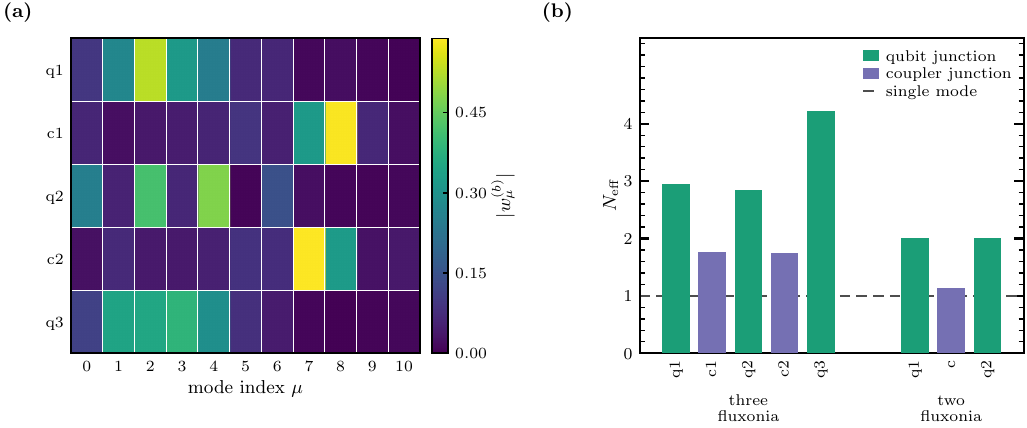}
\caption{Mode content of the two circuits, computed from the same normal
modes as Table~\ref{tab:modes}. (a) The phase participation
$|w^{(b)}_\mu| = |\alpha^{(b)}_\mu|\,\sigma_\mu$ of every Josephson junction of
the eleven-mode circuit, with
$\sigma_\mu = (8E_{C,\mu\mu}/E_{L,\mu\mu})^{1/4}/\sqrt2$ the zero-point
spread of $\hat\phi_\mu$, so that $w^{(b)}_\mu$ is the zero-point phase mode
$\mu$ contributes to junction $b$; junctions in chain order down the rows and
mode index across the columns, so that a row spread over several columns is a
cosine that cannot be assigned to one mode. (b) The participation ratio
$N_{\mathrm{eff}} = (\sum_\mu [w^{(b)}_\mu]^{2})^{2}/\sum_\mu [w^{(b)}_\mu]^{4}$ of every
junction of both circuits, where $N_{\mathrm{eff}} = 1$ would be a junction
that sits on a single mode. Junction labels are the subscripts of the Josephson
energies in Table~\ref{tab:params}.}
\label{fig:modes}
\end{figure*}

The normal-mode basis diagonalizes the linear circuit exactly.
Every junction except the coupler junction
of the two-fluxonium circuit spreads over several modes, as
Fig.~\ref{fig:modes} shows, and the cosine of junction $b$ becomes a product of
single-site operators over the support of $\alpha^{(b)}$. Panel (a) shows that support junction by junction and panel (b)
measures it, both through $w^{(b)}_\mu$. The scale of each column of $M$ is a
convention. Rescaling a column changes $E_{C,\mu\mu}$, $E_{L,\mu\mu}$ and
$\alpha^{(b)}_\mu$ but not $\Omega_\mu$ of Eq.~\eqref{eq:local} or the junction
terms of Eq.~\eqref{eq:junction-product}, which depend only on the product
$E_{C,\mu\mu}E_{L,\mu\mu}$ and on $\alpha^{(b)}_\mu\hat\phi_\mu$. The weight
$w^{(b)}_\mu = \alpha^{(b)}_\mu\sigma_\mu$, with $\sigma_\mu$ the zero-point
spread of $\hat\phi_\mu$, is therefore free of that convention. No junction of the three-fluxonium
circuit sits on one mode: its participation ratios $N_{\mathrm{eff}}$
run from \vEffModesThreeQ,
and the worst junction places only \vTopModeWeightWorstThreeQ{} of its
normalized weight on its own strongest mode. In the two-fluxonium circuit the
coupler junction sits almost on one mode
($N_{\mathrm{eff}} = \vEffModesTwoQCoupler$), while both qubit junctions sit
at \vEffModesTwoQQubit, a value set by the exchange symmetry of the two
identical qubits, which splits each junction evenly over the qubit mode pair,
and not by their coupling. Identical qubits spread their junctions over
several modes at any coupling strength, so the participation ratio does not
measure how large the coupling is. At two fluxonia the interaction itself
exceeds the qubits' own tunnel splitting where the tilt vanishes
(Sec.~\ref{sec:circuits}). The three-fluxonium circuit is characterized
structurally only.

The five low modes carry most of the qubit-junction weight, as panel (a) shows.
A hierarchical scheme that diagonalizes those five modes as one subsystem would
be a different method from the bare-basis HD of Sec.~\ref{sec:q2}, which keeps
one harmonic subsystem per mode, and we did not try it.

\section{Convergence criteria, schedule and status}
\label{app:convergence}

This appendix states the convergence protocol behind every tensor-network
number quoted in Secs.~\ref{sec:q2} and~\ref{sec:q3}: the conditions under
which a result is called converged, the DMRG schedule of each circuit, the
status of each circuit against those conditions, and the seeded continuation
that Sec.~\ref{sec:q2} reports, with one chain at three fluxonia.

Of the two truncation axes of Sec.~\ref{sec:truncation}, the bond cap truncates
the state: at a fixed Fock cutoff a matrix product state of bounded bond
dimension is a variational ansatz for the eigenstates of the truncated
Hamiltonian, so a bond ladder is read against its largest cap and its energy error is
one-sided, strictly for the ground state and to the accuracy of the
orthogonalization for the excited states.
The Fock cutoff truncates the operator,
and does so non-variationally: each junction factor in
Eq.~\eqref{eq:junction-product} is built as the exponential of the truncated
phase operator, not as the truncation of the exponential, so the Fock-truncated
Hamiltonian is not the projection of $H$ onto the kept levels, its error need
not fall monotonically with the cutoff, and a Fock ladder, a chosen path through
the vector of per-mode cutoffs, tests a plateau and never a bound. The two axes
also interfere, because a Fock ladder read at a binding bond cap measures the
ansatz rather than the operator, so each axis is swept with the other held at a
setting whose own error is below tolerance and is read against the largest
setting on the same axis.

A result is called converged only if all of the following hold at the quoted
setting. Every deviation is the metric of Sec.~\ref{sec:truncation}, the worst
transition deviation over all pairs of the quoted levels and the stated bias
grid, and is quoted per axis, never summed.

\begin{enumerate}
\item On every truncation axis, Fock and bond, each swept with the other held,
  each step up the ladder from the quoted setting moves the transitions by less
  than \vTolMHz{}. The error quoted for an axis is the distance to that axis's
  largest setting.
\item The bond axis satisfies one of two conditions: the achieved bond dimension
  lies below the cap at the discarded-weight cutoff \vCutoff, or the transitions
  agree to within \vTolMHz{} between the two largest caps with the largest at
  least two doublings above the quoted $\chi$. At eleven modes every solve
  reaches its cap, so only the second applies.
\item At least \vReplicaCount{} DMRG runs from independent random initial states
  agree on all quoted transitions to within \vTolMHz, at the working cutoff and
  at bond cap 64 for two fluxonia and 32 for three.
\end{enumerate}

Here $\chi$ is the bond dimension of Sec.~\ref{sec:truncation}, neither the
achieved bond dimension, which depends on the schedule, nor an entanglement
measure. The energy variance and the discarded weight of every solve are
recorded but are no convergence criterion, and neither is the worst top-Fock
occupancy,
the largest expectation value of the projector onto the highest kept level
over all modes and quoted states. It is reported beside each verdict and
flagged above \vTopFockCrit{}, but the stiff probe of Sec.~\ref{sec:q3-convergence}
exceeds the flag while moving the transitions by less than the tolerance.

Table~\ref{tab:dmrg} gives the DMRG schedule of each circuit. Every two- and
three-fluxonium solve follows it, except the seeded steps and the 16-level
rerun described below and the rerun of Sec.~\ref{sec:q2}.

\begin{table*}[tbp]
\centering
\caption{DMRG schedules of the two circuits. The bond and noise ramps are
applied sweep by sweep and then held. The local eigensolver is the library's
Hermitian Lanczos routine, one Krylov subspace per bond update and no restart,
so its dimension decides whether close levels are resolved: at $\Phi_c = 0$
and $\Phi_0/2$ the lowest two levels of the two-fluxonium circuit are nearly
degenerate, while the lowest eight levels of the three-fluxonium circuit lie at
least \vThreeQMinLevelSpacing{} apart. A few more states than quoted are solved
and the lowest are kept, so that a level converged out of order is not
missed.}
\label{tab:dmrg}
\begin{ruledtabular}
\begin{tabular}{lll}
Parameter & Two fluxonia & Three fluxonia \\
\colrule
Sweeps & \vSweepsTwoQ & \vSweeps \\
Bond-dimension ramp, clipped at the cap & \vMaxdimRampTwoQ & \vMaxdimRamp \\
Discarded-weight cutoff & \multicolumn{2}{l}{\vCutoff} \\
Noise ramp & \multicolumn{2}{l}{\vNoiseRamp} \\
Local Krylov dimension & \multicolumn{2}{l}{\vKrylovDim} \\
Orthogonality weight & \multicolumn{2}{l}{\vOrthoWeight} \\
States solved beyond the quoted levels & \vGuardStatesTwoQ & \vGuardStatesThreeQ \\
Initial bond dimension & \vInitBondTwoQ & \vInitBondThreeQ \\
Fock levels per mode & \vDimDCertTwoQ & \vDimDLowThreeQ{} on the low modes, \vDimDHighThreeQ{} on the others \\
Site order & \multicolumn{2}{l}{ascending $\omega_\mu$, fixed} \\
Start & \multicolumn{2}{l}{random state with a recorded seed} \\
\end{tabular}
\end{ruledtabular}
\end{table*}

\subsection{Status of each circuit}

At the \vGridPointsTwoQ{} grid points of Sec.~\ref{sec:q2}, at the working
cutoff and bond cap 64, the bond axis of the two-fluxonium circuit meets
criterion 2, with caps 64 and 128 as the two largest, and the replicas meet
criterion 3. The Fock axis, read along the ladder of \vDimDCertTwoQ, 16 and
\vFockFloorTopDim{} levels per mode, meets criterion 1 at
$\Phi_c = 0.4\,\Phi_0$ and misses it at three points, stated here in the
wording fixed before the runs. The plateau is missed at
$\Phi_c = 0.1\,\Phi_0$ by \vTwoQFockStepFirstPointTwo{} then
\vTwoQFockStepSecondPointTwo; the working cutoff stays \vDimDCertTwoQ. The
plateau is missed at $\Phi_c = 0.2\,\Phi_0$ by \vTwoQFockStepFirstPointThree{}
then \vTwoQFockStepSecondPointThree; the working cutoff stays \vDimDCertTwoQ.
The plateau is missed at $\Phi_c = 0.3\,\Phi_0$ by \vTwoQFockMissZThirtyA{} then
\vTwoQFockMissZThirtyB; the working cutoff stays \vDimDCertTwoQ. At
$\Phi_c = 0.3\,\Phi_0$ the 20-level solve lies \vTwoQFockTwentyVsFourteen{} from
the \vDimDCertTwoQ-level one, and the 16-level solve, rerun with \vTnProbeSweeps{} sweeps and
a local Krylov dimension of \vTnProbeKrylovDim{} and again from another random
state, is reproduced to \vTwoQFockSixteenRerun. At $\Phi_c = 0$ and $\Phi_0/2$,
where the lowest two levels are nearly degenerate, the Fock steps are of the
size of the replica spread, so the Fock axis is not resolved there. The worst
top-Fock occupancy at the working cutoff is \vTwoQTopFock, below the flag,
which only the cutoffs of 10 and 12 levels per mode exceed.

\begin{figure}[tbp]
\centering
\includegraphics[width=0.46\textwidth]{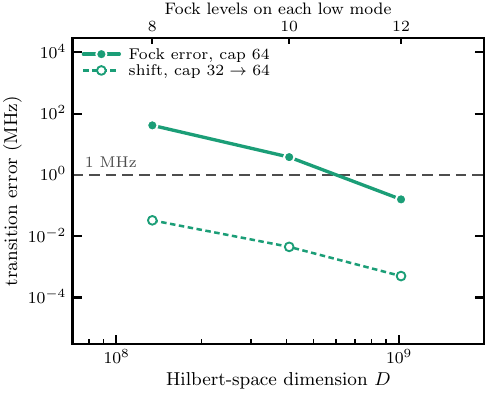}
\caption{Separating the two truncations on the eleven-mode Fock ladder of
Fig.~\ref{fig:q3}(b), for the lowest eight levels at the working point. The
solid line is the transition error of each cutoff against \vDimDRefThreeQ{} levels
on the low modes at bond cap 64 (top axis: Fock levels on each low mode); the
short-dashed line with open markers is how far that error moves, transition by transition, between caps
32 and 64. At every cutoff the shift lies far below both that cutoff's Fock error
and the \vTolMHz{} tolerance; at the working cutoff it is
\vThreeQBondMoveProd, the evidence for carrying the Fock error to cap 128.}
\label{fig:fock-bond}
\end{figure}

At three fluxonia every number of the convergence study is taken at the
working point
(Sec.~\ref{sec:q3-convergence}). The interference of the two axes was checked
there directly, as Fig.~\ref{fig:fock-bond} shows. The Fock ladder
of Fig.~\ref{fig:q3}(b) ran at bond caps 32 and 64, and at every cutoff below the
largest its Fock error was compared between the two caps transition by
transition, keeping the largest shift. Differencing only the worst transition
error of each cap would miss a Fock error that moves to another transition when
the cap changes. The quoted transitions are those of the cap-128 solve, and no
Fock ladder was run at that cap: the Fock error measured at caps 32 and 64 is
carried to cap 128, an assumption supported by the shift of that error,
\vThreeQBondMoveProd, between those two caps at the working cutoff.

Table~\ref{tab:q3levels} lists the seven transitions of the lowest eight levels
with the move along each axis. The solves at the working cutoff were repeated in
processes of their own, which reproduce the levels to \vThreeQBridgeMax; the
other solves of the convergence study ran once.

\begin{table}[tbp]
\centering
\caption{The transitions from the ground state of the lowest eight levels of
the three-fluxonium circuit at the working point, from the cap-128 solve at
the working cutoff, with the move along each axis. The Fock step is the move from
\vDimDLowThreeQ{} to \vDimDRefThreeQ{} levels on the low modes at cap 64,
carried to cap 128; the stiff shift is the move when the four stiff modes are
raised to \vDimDStiffProbeThreeQ{} levels; the bond step is the move from cap 64
to 128. The levels are named by count and not shown to be the eight states of
three qubits; they lie at least \vThreeQMinLevelSpacing{} apart.}
\label{tab:q3levels}
\begin{ruledtabular}
\begin{tabular}{lrrrr}
Level & Transition & Fock step & Stiff shift & Bond step \\
\colrule
1 & \vThreeQTransitionOne & \vThreeQFockBarOne & \vThreeQStiffShiftOne & \vThreeQBondBarOne \\
2 & \vThreeQTransitionTwo & \vThreeQFockBarTwo & \vThreeQStiffShiftTwo & \vThreeQBondBarTwo \\
3 & \vThreeQTransitionThree & \vThreeQFockBarThree & \vThreeQStiffShiftThree & \vThreeQBondBarThree \\
4 & \vThreeQTransitionFour & \vThreeQFockBarFour & \vThreeQStiffShiftFour & \vThreeQBondBarFour \\
5 & \vThreeQTransitionFive & \vThreeQFockBarFive & \vThreeQStiffShiftFive & \vThreeQBondBarFive \\
6 & \vThreeQTransitionSix & \vThreeQFockBarSix & \vThreeQStiffShiftSix & \vThreeQBondBarSix \\
7 & \vThreeQTransitionSeven & \vThreeQFockBarSeven & \vThreeQStiffShiftSeven & \vThreeQBondBarSeven \\
\end{tabular}
\end{ruledtabular}
\end{table}

\subsection{Seeded continuation}
\label{app:seeding}

The seeding of Sec.~\ref{sec:q2} ran one chain per bond cap along the
two-fluxonium grid. A chain starts with a cold solve on the schedule of
Table~\ref{tab:dmrg}; each later point starts from the states converged at the
previous one, at the same Fock cutoff, cap and mode basis, and runs
\vWarmSweeps{} sweeps at a flat cap with one noise sweep at \vWarmNoise. Five
checks, fixed before the runs, decide whether a seeded step stands, and a step
that fails one is re-solved cold and starts a new chain. The states must stay
orthogonal, with no overlap above \vWarmOrthoTol; no energy variance may exceed
\vWarmVarianceFactor{} times the worst of the chain's cold start, as a check
that every state is still an eigenstate; no energy may move by more than
\vWarmSettleTol{} over the last two sweeps, as a check that the step has
settled; the level order must not
change; and the step must finish within the wall time of the chain's cold
start.

In all, \vWarmStepsPassed{} seeded steps passed. Of the \vWarmStepsFailed{} that
failed, \vWarmCrossingOnly{} failed only the level-order check and the others
at least the settling check; none exceeded the time budget. With every attempt
counted and the
wall time summed over grid points 2 to 6, the cold solves of
Sec.~\ref{sec:q2} took \vWarmSpeedupCapEight, \vWarmSpeedupCapSixteen{} and
\vWarmSpeedupCapThirtyTwo{} times the wall time of the chains at bond caps 8,
16 and 32, and \vWarmSpeedupCapSixtyFour{} times at cap 64, on the same host.
For comparison, repeating the cold solve of one grid point took
\vColdResolveSpread{} times its wall time in Sec.~\ref{sec:q2}. A passing step ran \vWarmStepSpeedup{} times faster than the
cold solve of its point, close to the ratio of the sweep budgets,
\vWarmSweeps{} against \vSweepsTwoQ. The seeded solves that passed lie within
\vWarmColdDevMax{} of the cold ones but farther from the exact reference: at
$\Phi_c = 0.3\,\Phi_0$ and bond caps 16 to 64 they deviate from it by
\vWarmVsRefSeeded, the cold solves by \vWarmVsRefCold.

At three fluxonia one chain ran at \vDimDBareRefThreeQ{} Fock levels on each low
mode and bond cap 32, starting at the working point and stepping the flux of
the first coupler by $\Phi_0/32$ up to $\Phi_0/8$ and back.
Of its eight seeded steps, \vThreeQWarmFailedSteps{} failed and were re-solved
cold, and the seeded return to the working point agrees with the cold solve
there to \vThreeQWarmReturnMHz. No speedup is quoted for this
chain.

\section{Computing environment}
\label{app:hardware}

Every wall time and peak memory quoted in this work was measured on one CPU node with
an \vCpuModel{} processor, \vBenchThreads{} physical cores without simultaneous
multithreading and \vNodeMem{} of RAM, while no other job ran on it. The timed
solvers ran in Julia~\vJuliaVersion~\cite{bezanson_julia_2017} with
ITensors~\vITensorsVersion, ITensorMPS~\vITensorMPSVersion~\cite{itensor} and
KrylovKit~\vKrylovKitVersion~\cite{haegeman_krylovkit}. Exact HD of
Sec.~\ref{sec:q2} applies our own matrix-free Hamiltonian in KrylovKit's Lanczos
eigensolver, with \vEdKrylovVectorsTwoQ{} Krylov vectors, a residual tolerance
of \vKrylovTolTwoQ, threads tuned on the CPU node and a seeded random start.

\bibliographystyle{apsrev4-2}
\bibliography{refs}

\end{document}